%% file: main.tex
\documentclass[conference,compsoc]{IEEEtran}

\ifCLASSOPTIONcompsoc
  \usepackage[nocompress]{cite}
\else
  \usepackage{cite}
\fi

\input{packages}

\begin{document}

\title{Towards Automating Data Access Permissions in AI Agents}

\author{%
\IEEEauthorblockN{%
    Yuhao Wu\IEEEauthorrefmark{1}, 
    Ke Yang\IEEEauthorrefmark{2}, 
    Franziska Roesner\IEEEauthorrefmark{3},
    Tadayoshi Kohno\IEEEauthorrefmark{4}, 
    Ning Zhang\IEEEauthorrefmark{1}, 
    Umar Iqbal\IEEEauthorrefmark{1}%
}
\IEEEauthorblockA{\IEEEauthorrefmark{1}Washington University in St. Louis\,
                  \IEEEauthorrefmark{2}University of California, Irvine\,
                  \IEEEauthorrefmark{3}University of Washington\,
                  \IEEEauthorrefmark{4}Georgetown University}
}

\maketitle
\thispagestyle{plain}
\pagestyle{plain}

\input{0_abstract}

\input{1_introduction}

\input{2_motivation}

\input{3_threat_model}

\input{4_understanding_preferences}

\input{5_predicting_preferences}

\input{6_discussion}

\input{7_conclusion}

\input{acknowledgment}

\input{references.bbl}
\appendices

\input{X_appendix}

\newpage %

\input{X_meta_review}

\end{document}

%% file: packages.tex
\usepackage{graphicx}
\usepackage{subcaption}
\usepackage{multicol}
\usepackage{multirow}
\usepackage{float}
\usepackage{caption}
\usepackage[table]{xcolor}
\usepackage{booktabs}
\usepackage{pgfmath}
\usepackage{listings}
\usepackage{longtable, array}
\usepackage{hyperref}
\usepackage{enumitem}
\usepackage{todonotes}

\definecolor{gheat1}{RGB}{229,245,224}  %
\definecolor{gheat2}{RGB}{199,233,192}
\definecolor{gheat3}{RGB}{161,217,155}
\definecolor{gheat4}{RGB}{116,196,118}
\definecolor{gheat5}{RGB}{65,171,93}
\definecolor{gheat6}{RGB}{35,139,69}
\definecolor{gheat7}{RGB}{0,109,44}     %

\newcommand{\gheatbg}[1]{%
  \pgfmathparse{#1}%
  \ifdim\pgfmathresult pt<25pt \cellcolor{gheat1}#1\%%
  \else\ifdim\pgfmathresult pt<35pt \cellcolor{gheat2}#1\%%
  \else\ifdim\pgfmathresult pt<45pt \cellcolor{gheat3}#1\%%
  \else\ifdim\pgfmathresult pt<55pt \cellcolor{gheat4}#1\%%
  \else\ifdim\pgfmathresult pt<60pt \cellcolor{gheat5}#1\%%
  \else\ifdim\pgfmathresult pt<65pt \cellcolor{gheat6}#1\%%
  \else \cellcolor{gheat7}#1\%%
  \fi\fi\fi\fi\fi\fi
}

\definecolor{rheat1}{RGB}{254,229,217}
\definecolor{rheat2}{RGB}{252,187,161}
\definecolor{rheat3}{RGB}{252,146,114}
\definecolor{rheat4}{RGB}{251,106,74}
\definecolor{rheat5}{RGB}{239,59,44}
\definecolor{rheat6}{RGB}{203,24,29}
\definecolor{rheat7}{RGB}{153,0,13}     %

\newcommand{\rheatbg}[1]{%
  \pgfmathparse{#1}%
  \ifdim\pgfmathresult pt<4pt \cellcolor{rheat1}#1\%%
  \else\ifdim\pgfmathresult pt<7pt \cellcolor{rheat2}#1\%%
  \else\ifdim\pgfmathresult pt<10pt \cellcolor{rheat3}#1\%%
  \else\ifdim\pgfmathresult pt<13pt \cellcolor{rheat4}#1\%%
  \else\ifdim\pgfmathresult pt<15pt \cellcolor{rheat5}#1\%%
  \else\ifdim\pgfmathresult pt<17pt \cellcolor{rheat6}#1\%%
  \else \cellcolor{rheat7}#1\%%
  \fi\fi\fi\fi\fi\fi
}

\usepackage{pgf} %

\usepackage{xcolor} %

%% file: 0_abstract.tex
\begin{abstract}
As AI agents attempt to autonomously act on users' behalf, they raise transparency and control issues. 
We argue that permission-based access control is indispensable in providing meaningful control to the users, but conventional permission models are inadequate for the automated agentic execution paradigm. 
We therefore propose \textit{automated permission management for AI agents}. 
Our key idea is to conduct a user study to identify the factors influencing users' permission decisions and to encode these factors into an ML-based permission management assistant capable of predicting users' future decisions.
We find that participants' permission decisions are influenced by communication context but importantly individual preferences tend to remain consistent within contexts, and align with those of other participants. %
Leveraging these insights, we develop a permission prediction model achieving 85.1\% accuracy overall and 94.4\% for high-confidence predictions.
We find that even without using permission history, our model achieves an accuracy of 66.9\%, and a slight increase of training samples (i.e., 1--4) can substantially increase the accuracy by 10.8\%.

\end{abstract}

%% file: 1_introduction.tex
\section{Introduction}
LLMs have enabled a new computing paradigm, in which the system (referred to as an \textit{AI agent} or an \textit{agentic system}) relies on machine learning models to autonomously resolve user queries expressed in natural language~\cite{wiesinger2025agents}.
For example, to resolve a user query to \textit{book a flight}, the AI agent might engage with the necessary tools (e.g., a travel reservation tool), automatically access the required user information (e.g., from the system storage/memory), use the user's credit card details, and make the booking.
While this execution paradigm is tremendously powerful and is enabling exciting use cases~\cite{openai_operator_2025, google_project_mariner_2024, anthropic_computer_use_2024}, there are serious security and privacy risks to consider.

At a high level, a key concern is that the AI agent's actions may not align with the user's intentions or expectations, which could be due to untrustworthy system modules influencing the LLM~\cite{iqbal2024llm,jaff2024data,zhao2024llm}, LLM making mistakes on its own~\cite{liu2023we,yao2024survey,huang2025survey}, or the user simply not agreeing with the agent's course of action~\cite{wang2024ali,li2024dissecting,movva2024annotation}.
These concerns apply to a broad range of the agent's activities, such as autonomous data access and actions in the real-world.
In this paper, \textit{we focus on the security and privacy issues pertaining to the access of data}.
For example, we attempt to limit the unexpected sharing of a user's credit card details with the wrong tool.

Fundamentally, the key issues are limited user transparency and control over the data access by AI agents.
Prior computing systems, such as mobile platforms, have encountered similar problems and have relied on permission models (in which users are asked to grant applications permission to access sensitive resources) to provide users transparency and control over data usage in the system~\cite{roesner2012user,ringer2016audacious,harbach2024don,felt2012android,felt2011android}.
Similarly, AI agents will need permission management models to control access to resources.

Permission management, however, needs to be designed anew for AI agents, where new challenges arise that stretch the limits of existing permission models. %
For example, as AI agents generate new functionalities based on input from various system modules, data needed to resolve user queries may not be known beforehand, thus diminishing the utility of legacy install time permissions. %
Similarly, as AI agents may need several pieces of user information to resolve a user query, constantly interrupting users for permissions is incompatible with the automation promised by AI agents, thus diminishing the utility of standard runtime permissions.

Considering that \textit{automation} is a key value proposition of AI agents, \textit{a permission management system that can automatically make decisions on users' behalf is critical for AI agents}.
Building an automated permission management requires addressing two important problems.
First, understanding the factors that users consider or factors that otherwise influence users' permission decision-making. 
Second, making permission decisions that precisely meet user needs and expectations for future data sharing, including previously unseen data types.
This paper takes a multifaceted approach, including both \textit{(i)} conducting a user study to understand user permission preferences and \textit{(ii)} exploring the design of a permission prediction system along with its implementation and evaluation.

To understand users' data sharing permission preferences, we conduct a vignette-based user study. 
We create a bespoke user study setup by developing our own website that attempts to immerse participants in training a futuristic personal assistant, including training it to share data, as per their needs. 
We capture differing user preferences across a wide range of questions spanning several domains, such as health and fitness, finance, and entertainment. 
Our findings indicate that fewer participants express permissions to AI agents for automatic enforcement when they make mistakes, and participants often struggle with providing appropriate permissions, i.e., under- and over-permissioning are common issues in agentic systems, similar to prior systems~\cite{seneviratne2015your,peddinti2019reducing,lin2014modeling}. %
We also note that participants' permission decisions are influenced by the context of the communication, choice of data being shared, and their privacy consciousness.  
Importantly, we find that, at least in the context of our study, participants' permission preferences remain consistent, are similar across various communication contexts, and are often similar to other participants---presenting an opportunity to predict their future permissions.

To learn and predict users' data sharing permission preferences, we explore developing a hybrid machine learning framework that learn from individual user preferences and also from preferences of similar users (using data from our user study). %
We leverage LLM in-context learning~\cite{nori2023can} to learn individual user preferences for two crucial reasons: \textit{(i)} we possess limited permission decision history for each user, and LLMs have demonstrated attaining high accuracy with limited training data, i.e., ``few shots''~\cite{nori2023can,dong2024survey}; \textit{(ii)} permission decisions need to be made for unseen data, which LLMs allow to make without retraining~\cite{brown2020language}.
We leverage collaborative filtering~\cite{he2017neural,he2020lightgcn} because it allows us to learn from the preferences of other similar users. 
Our final resulting model combines both in-content learning and collaborative filtering, as they complement each other. 
We achieve an accuracy of 85.1\% (with a recall of 85.2\% and a precision of 92.8\%). 
We also explore adjusting prediction confidence score thresholds and find that a stricter threshold, we can achieve an accuracy of 94.4\%, but compromise on making predictions for 74.1\% of the data.  %
We also observe that even a slight increase in training data (i.e., user permission decision history), our classification accuracy substantially increases. %
For example, accuracy reaches 66.9\% without permission history, but incorporating history from just 1–4 queries improves it by 10.8\%.
To foster future research, we release our user study data and code\footnote{\url{https://github.com/llm-platform-security/ai-agent-permissions}}.

Our key contributions are as follows:
\begin{enumerate}[leftmargin=5mm]
    \item We propose automating data access permissions in AI agents, such that a \textit{permission assistant} can observe a user's permission decision history and can make automatic decisions on the user's behalf in the future.

    \item To realize our goal of automating permission decisions, we develop a bespoke vignette-based user study to understand various factors that may influence users' data-sharing permission decisions in AI agents. We then conduct the study with 205 participants on Prolific. 

    \item We translate the insights from our user study into a permission inference framework capable of predicting users' permission preferences, achieving 85.1\% accuracy overall and 94.4\% for high-confidence predictions.

\end{enumerate}

%% file: 2_motivation.tex
\section{Background and Motivation}

\subsection{AI Agents}
While there is no standardized AI agent architecture, at their core, AI agents (often also referred to as \textit{agentic systems}) consist of an LLM (typically accessed via an API), a system prompt (that defines the agent's functionality), memory (to keep a record of user interactions and data), and a set of tools (to take action in the real world)~\cite{wiesinger2025agents}. 
To resolve a user query, AI agents identify the relevant tools and data, formulate an execution plan (i.e., a set of instructions) for an LLM, and autonomously act on the formulated plan.

For example, for a user query to ``book a flight'', the agent will first determine that it needs to use a travel reservation tool and requires the user's information (such as the user's name and date of birth).
The execution plan may include instructions to search for flights, provide payment details, and make the booking. 
Acting on the plan will include the agent calling the appropriate travel reservation tool APIs with the relevant data, listed in the execution plan.

AI agents exhibit varying degrees of autonomy; in some cases, they can complete tasks entirely of their own, while in other cases they require some level of human supervision during execution.
For example, for the flight booking request, agents may retrieve trip dates and locations from the user query, the user's name and date of birth from the memory storage, and may only request the user to provide credit card information.

\subsection{Security and Privacy Risks}
\label{sec:sec-pri-risks}

As the automated execution paradigm of AI agents makes user interactions seamless, there are obvious benefits to it; however, it also presents serious security, privacy, and safety issues to the users. 
At a high level, a key concern is that the AI agent's actions may not align with the user's intentions or expectations.
The misalignment could be due to a variety of reasons, such as untrustworthy system modules influencing the LLM, LLM making mistakes of its own, or the user simply not agreeing with the agent's course of action. 
These concerns apply to a broad range of agents' activities, such as agents using user data of their own or taking actions on users' behalf in the real world, which are distinct and require tailored treatment. 
For the scope of this paper, we only focus on limiting the security and privacy issues that pertain to the usage of data.

Next, we describe some of the fundamental issues that could lead to a misalignment between users' expectations and AI agents' data usage practices.

\vspace{1mm}
\noindent
\textbf{Inherent Limitations in Agentic Execution Paradigm.}
AI agents enhance their capabilities through exposure to system resources, such as data and tools. 
It means that only when an agent is aware of the system capabilities (i.e., a particular tool or a piece of data) can it use those. 
Thus, agents are often designed to get unrestrained access to system resources, including the ones that may not be needed to address the user query, which violates the principle of least privilege. %
As LLMs are susceptible to prompt injection, malicious resources (such as malicious third-party tools or files) can exploit the LLM's unrestrained system access to read sensitive user data or influence the LLM's execution. %

Another key issue is that the LLM's interfacing is based on natural language, which can be imprecise and ambiguous~\cite{liu2023LLMAmbguity, iqbal2024llmAIES}. 
It means that even in non-adversarial scenarios, the underlying LLM in an AI agent might collect incorrect or unnecessary user data. 
For example, if the travel reservation tool imprecisely specifies that it needs \textit{relevant user data} to make a travel reservation, the LLM's interpretation of this data may be different than that of the tool and result in inadvertent collection and sharing of unnecessary user data. 
Hallucination issues in LLMs may further exacerbate these problems~\cite{iqbal2024llmAIES}.

\vspace{1mm}
\noindent
\textbf{Untrustworthy Third-party Tools.}
AI agents rely on input from several system modules to determine and steer their execution. 
Some of these modules, chiefly tools, are developed by third-party developers and load unvetted content from the internet, which makes them an unreliable source to determine the system's execution. 
For example, a malicious, compromised, or buggy travel reservation tool may direct the agent to include users' passport numbers for booking flights even when it is not necessary (such as for domestic flights). 
Prior research has already shown that third-party tools often collect more data than is needed, including sensitive data prohibited by the platforms~\cite{jaff2024data}.
While the problem of excessive data collection by third-party services has existed in prior computing platforms~\cite{EnglehardtOpenWPMMeasurement,razaghpanah2018apps,iqbal2022your,trimananda2022ovrseen}, it presents elevated risks in the case of AI agents, because of their pivotal reliance on third-party tools to determine their execution.

\vspace{1mm}
\noindent
\textbf{Users' Privacy Consciousness.}
Users' expectations, desires, and privacy needs may also simply not align with an AI agent's action.
For example, prior research has shown that users often feel uneasy about sharing intimate data (e.g., health, finances) with AI assistants, fearing misuse or eavesdropping~\cite{malkin2022runtime,lau2018alexa}.
Prior research also shows that users differ significantly in their willingness to share data, which is often influenced by context trust, tech literacy, and prior experience~\cite{abdi2021privacy,malkin2022runtime,bagdasarian2024airgapagent}.
In some cases, users may refuse to avail some services or compromise on the user experience. 
For example, in the context of the flight booking example, users may prefer providing the origin flight city manually, instead of having an agent infer it from the device's GPS sensor.

\subsection{Permission-Based Data Access}
\label{subsection:Permission-Based-Data-Access}
Fundamentally, the key issues are limited transparency and control over the data access by AI agents. 
Prior computing systems and platforms, such as mobile platforms, have encountered similar problems and have relied on \textit{permission models} to provide user transparency and control over the data usage in the system~\cite{roesner2012user,ringer2016audacious,harbach2024don,felt2012android,felt2011android}.
Likewise, agents can also benefit from a permission management model to control access to resources.

Deployed AI agents, such as ChatGPT, currently adapt permission models from conventional systems~\cite{openai_consequential_flag}; however, as we describe next, they fall short in supporting and instead hinder the automated agentic execution paradigm. %

\vspace{1mm}
\noindent
\textbf{Insufficiency of Existing Permission Models.}
Permission management systems for AI agents need to be developed anew as they significantly differ from conventional computing systems. 
Specifically, the install-time~\cite{AndroidPermissions} and runtime~\cite{AndroidRuntimePerms} permissions from conventional systems are insufficient, especially considering the automated execution paradigm of AI agents. 
For example, install-time permissions assume that the resources needed by applications are known beforehand, and thus users can make permission decisions at the time of installation. 
In contrast, in most cases of agent execution, behavior is determined at runtime based on input from system modules, which can lead to the emergence of new behaviors.
Thus, the resources (which can be very broad) needed to solve a query may not be known beforehand~\cite{wiesinger2025agents,yao2023react}.
Similarly, runtime permissions (upon which smartphone platforms have largely converged) allow users to manually make decisions on access of individual resources, which suits conventional systems as only a handful of resources are accessed at runtime (and where user interactions already follow a typical UI flow), such as granting location access permission in mobile platforms.
Whereas agents often require accessing several pieces of user data at runtime to solve a query, thus merely applying legacy runtime permission models can degrade the user experience and contribute to permission fatigue.

While conventional permission models are mostly limited in supporting the agentic execution paradigm, they could still be suitable for some use cases. 
For example, AI agents could leverage install-time permissions to manage OAuth-based authentication~\cite{openai-gpt-action-authentication} in AI agents.

%% file: 3_threat_model.tex
\section{Towards Automated Permissions Management in AI Agents}

Considering that a significant number of data resources are accessed at runtime to facilitate execution of queries---more than users can reasonably be asked to constantly evaluate and decide upon---we argue that a permission management system that can automatically make decisions on users' behalf is a necessity for agents. 
Our observation is consistent with previous work on voice-based personal assistants, which found that while users want control, they dislike excessive prompts and prefer automated permission management with minimal interruptions~\cite{malkin2022runtime}, and permission prompt fatigue has been a longstanding known issue in other contexts as well~\cite{shao2022you,felt2012android,bilogrevic2021shhh,motiee2010windows}.

\subsection{Research Goals}
Developing automated permission management requires addressing three key challenges: (i) understanding diverse user data sharing preferences, (ii) accurately learning and predicting user preferences, and (iii) reliably enforcing predicted preferences. 
In this paper, we focus on (i) and (ii); prior complementary work~\cite{bagdasarian2024airgapagent,wu2025isolategpt,Moskal2024} can be used to address (iii).
We describe our approaches to achieve these goals below. 

\subsubsection{Goal 1: Understanding diverse user preferences}
A prerequisite to making permission decisions on users' behalf is to understand the preferences and expectations of a variety of users. 
To understand user preferences, we conduct a vignette-based user study, which presents several scenarios to participants to capture their preferences.
Building on the prior work, we identify several factors, such as the context of the communication, privacy consciousness, and users' prior experiences, that may influence user preferences~\cite{malkin2022runtime,lau2018alexa,abdi2021privacy,bagdasarian2024airgapagent}.
Our user-study presents several scenarios to users to capture their preferences across a wide range of factors that may influence their preferences. 

While prior work exists on understanding user preferences and expectations for personal assistants, it mostly focused on older non-LLM technologies~\cite{abdi2021privacy,malkin2022runtime,liu2022effects,malkin2019privacy,lau2018alexa}.
Newer LLM-based personal assistants/agents are fundamentally different as they rely on a new natural language-based automated execution paradigm and offer far more advanced capabilities. %
For example, non-LLM agents mostly supported single-dialog user interactions and carried out limited tasks automatically; consequently, prior studies only analyzed a handful of scenarios while examining user permission preferences~\cite{malkin2022runtime}.
Whereas LLM-agents support multi-dialog user interactions and carry out many tasks automatically~\cite{manusai, yao2023react}. 
Users are likely to form different mental models for modern LLM-based AI agents than for traditional, non-LLM agents, so tailored user studies are essential to capture these distinct experiences.
To the best of our knowledge, we are the first to study user preferences in automating data sharing permissions in the context of LLM-based AI agents.  
Our goal is also not just to simply understand user preferences, but to translate them to a system design and explore their utility in automatically predicting permission decisions. 
Prior work has also not explored understanding user permission preferences in AI-based systems for the purposes of training an automatic permission prediction assistant.

\subsubsection{Goal 2: Accurately learning user preferences}
To enforce user preferences across a wide spectrum of conversational contexts and a range of data types, it is crucial to develop a permission preference prediction system that accurately learn users' preferences from a handful of contexts and applies to other unseen contexts. 
More specifically, in a real world setting, users may only provide preferences for a select few scenarios, and the system may encounter scenarios that it has not seen before. 
To tackle these challenges, we rely on a combination of collaborative filtering~\cite{he2020lightgcn} and LLM-based in-context learning~\cite{nori2023can} to develop our permission inference system. 
We choose collaborative filtering because it enables learning a user's preferences by analyzing the preferences of similar users~\cite{he2017neural,he2020lightgcn}, thus avoiding the data sparsity issues. 
We choose LLM in-context learning, as it can allow the systems to continuously refine permission inference and predict new previously unseen data types, without requiring re-training, as new contexts and data types are seen~\cite{nori2023can,dong2024survey}.

Prior work has proposed predicting user permissions in the context of mobile~\cite{liu2016follow,filipczuk2022automated,brandao2022prediction,mendes2022enhancing, xie2014location}, IoT~\cite{das2018personalized,barbosa2019if,shanmugarasa2021automated}, and non-LLM personal assistants~\cite{zhan2022model,amoros2023predicting,zhan2023privacy}. %
However, prior work can only predict a handful of \textit{known} system resources and data types, which suited the needs of older systems.
For instance, notable studies on non-LLM agents/assistants manage data access permissions across only 15 data types at a coarse granularity~\cite{zhan2022model,amoros2023predicting,zhan2023privacy}.
In contrast, LLM-based AI agents routinely encounter new previously unseen data types as users explore new use cases~\cite{jaff2024data}, and thus prior approaches simply cannot scale to AI agents.
For example, a classic ML model trained on a set of data types used by a tool/app would require retraining to make predictions for new unseen data types used by the tool/app.
Our proposed approach tackles this challenge by relying on LLM in-context learning, which does not require retraining to make permission decisions on unseen data.

\subsection{Threat Model}

\noindent
\textbf{System Model.}
We assume AI agents that maintain user collected data in dedicated \textit{memory modules} and attempt to automatically use that data to provide personalized services to the users. 
These AI agents include well-known personal assistants, such as OpenAI's ChatGPT~\cite{openai_chatgpt}, as well as AI agents developed through agent development toolkits, such as LangChain~\cite{langchain} and LlamaIndex~\cite{llamaindex}.
These AI agents also support third-party tools, and both the AI agent and third-party tools can collect and use user data. 
AI agents also include system scaffolding that allows them to control the execution flow, e.g., accessing data from memory, initiating LLMs, and making network requests via tools~\cite{wiesinger2025agents}.  

\vspace{2mm}
\noindent
\textbf{Adversaries and Goals.}
We assume that the third-party tools could be untrustworthy/dishonest, malicious, or compromised (e.g., via a prompt injection). 
The attacker's goals are to leverage third-party tools to steal sensitive data that is present in the agent's memory or exists with other tools.
We assume the LLM to be not malicious but error-prone and making mistakes in accessing unnecessary and sensitive data, e.g., due to ambiguity of natural language~\cite{liu2023LLMAmbguity,iqbal2024llmAIES}.

\vspace{2mm}
\noindent
\textbf{Trust Relationships.}
We assume that the AI agent and its scaffolding to be trustworthy, and do not have any direct intent to harm users (though they are still vulnerable to attacks, e.g., prompt injection).
We also assume that the users may not share data with the AI agents and tools when they think it is unnecessary or simply do not feel comfortable sharing that data, despite data being necessary for the AI agents and tools to provide functionality. 

\vspace{2mm}
\noindent
\textbf{Permission Assistant Design Assumptions.}
To understand user preferences, we communicate to users (in our user study) that the tools may be malicious, compromised, or dishonestly collect more data than they need. 
We envision the deployment of our permission assistant in the system scaffolding of the AI agent, without the LLM having direct access to the permission assistant. 
As the permission assistant makes probabilistic decisions, it may unintentionally make incorrect predictions in some cases.
However, the permission assistant does not act with malicious intent, and users retain control over whether to accept its decisions.
Based on these considerations, we assume the permission assistant to be trustworthy.

\noindent
\textit{Out of scope.}
As the implementation and deployment of the permission assistant requires solving unique challenges of its own, we do not consider it in the scope of this paper. 
For example, a secure personal assistant may require sandboxed or TEE-based deployment, detection of malicious data flows, and control of the generation of LLM instructions. 
We envision that the prior work on these topics in the context of AI-agents (e.g.,~\cite{bagdasarian2024airgapagent,wu2025isolategpt,Moskal2024,debenedetti2025defeating}) can be extended to implement a secure permission assistant.

%% file: 4_understanding_preferences.tex
\section{Understanding User Preferences}
To automatically make permission decisions on users' behalf, it is crucial to first understand how users make permission decisions in the context of agentic systems. 
To this end, we conduct a user study to understand users' permission decision process. 
Our goal with the user study is to understand factors that influence users' decision making, so that they can help inform the design of automated permission management systems.

\subsection{Study Design}
\label{subsection:study-design}

\subsubsection{Overview}
We consider several variables in our user study (such as demographics, privacy consciousness, usage context) that prior research has identified to influence users' permission decisions in other computing platforms~\cite{malkin2022runtime,abdi2021privacy,bagdasarian2024airgapagent}.
Meanwhile, AI agents introduce unique capabilities that users have not experienced in prior systems. To help users become familiar with these capabilities and their associated risks, and to examine the influence of all factors in a controlled setting, we develop a vignette-based study~\cite{finch1987vignette}, which has been used in prior work investigating users' privacy preferences~\cite{lassakbalancing,lassakintroducing,schwab2025makes,li2022scenario,harborth2021investigating,naeini2017privacy}.
In our study, we present vignettes to users that attempt to immerse the participants in training a futuristic personal assistant to their needs. 
Specifically, we present scenarios to participants, where they are asked to help the personal assistant: (i) pick the right set of tools to solve a query, (ii) pick the right set of data to solve a query, and (iii) automatically make data accessing and sharing permission decisions on their behalf. %
We present a total of 5 questions for tool and data selection, and 20 questions for permission decisions.\footnote{We present 4 options for permission decisions: (1) Yes, always share,  (2) Yes, but ask me next time, (3) No, but ask me next time, and (4) No, never share. These options are the same as the permission options used by OpenAI's custom GPTs~\cite{chatgpt_gpts_announcement} (except for never share, which is not an option in OpenAI's ecosystem).}
To investigate how potential adversarial manipulations or mistakes may influence users' decisions, for 25\% (5) of permission decision questions, we include incorrect (unnecessary) data types and ask participants to express their data sharing permission preferences.
(Our analyses in Section~\ref{subsubsection:delegation} and~\ref{subsubsection:permission-practices} consider both incorrect and correct data, while the remaining sections only consider correct data.) %
We also include a warning for all permission decision questions that prompts the participants to be careful, as the agent or tools may collect incorrect or unnecessary data.
Additionally, we develop our own custom website to make the experience more immersive for users\footnote{We present the screenshots of our user study website at \url{https://github.com/llm-platform-security/ai-agent-permissions/blob/main/website.pdf}.}.

\subsubsection{Question curation}\label{subsubsection:query-curation}
We develop an LLM-based framework to curate questions for our user study. 
We begin by selecting a set of 8 domains\footnote{Entertainment, Health \& Fitness, Smart Home, Travel, Shopping, Work \& Productivity, Social, and Finance.} and curating 21 tools (spanning all domains) to record participant preferences on a variety of topics. 
Our tool curation involves listing functionalities offered by the tools and the data needed to provide those functionalities, leveraging prior work on tool curation~\cite{jaff2024data,debenedetti2024agentdojo}.
We treat the data needed by tools as ground truth.

We iteratively provide domains and their tools to the LLM, and prompt it to generate queries that users might ask an AI agent, and that require using one or more tools.
Once queries are generated, three members of our research team review the generated queries to filter out redundant queries and semantically group similar data types (e.g., address and location).
In summary, we curate 65 questions\footnote{
The questions are detailed at \url{https://github.com/llm-platform-security/ai-agent-permissions/blob/main/queries.json}.} spanning 8 domains, involving 142 unique data types and 75 generic data types across 21 tools.

\subsubsection{Priming participants}
Before participants pursue the questions in the user study, we prime the participants by asking them questions about their privacy consciousness (e.g., the importance of privacy to the participants, participants' key privacy concerns). %
Prior research has shown that asking about privacy attitudes at the start of a study can raise participants' privacy awareness and prompt more cautious behavior throughout the task~\cite{reis2000handbook,sotirakopoulos2011challenges,distler2023empirical}.
Our goal with priming is to make participants privacy conscious, so that their choices more accurately reflect their privacy posture, as it may in real life when they are training an assistant to make permission decisions on their behalf. %

\subsubsection{Participant recruitment}
We recruit a total of 205 participants (we remove 2 participants because of invalid responses) from Prolific~\cite{prolific2024} in the US, with a minimum Prolific prior survey approval rate of 90\% and age over 18 years. 
Our study takes approximately 18 minutes to complete, and we pay \$4.20  to each participant (using \$14 as the hourly wage at the lead institution).
Our study was reviewed by our institution's review board (IRB) and deemed exempt.

\subsection{Findings}
\label{sec:preferences_findings}

\subsubsection{When AI agents make mistakes, fewer participants express data sharing permissions, but more participants express not sharing permissions}
\label{subsubsection:delegation}
\begin{figure}[t]
    \centering
    \includegraphics[width=0.4\textwidth]{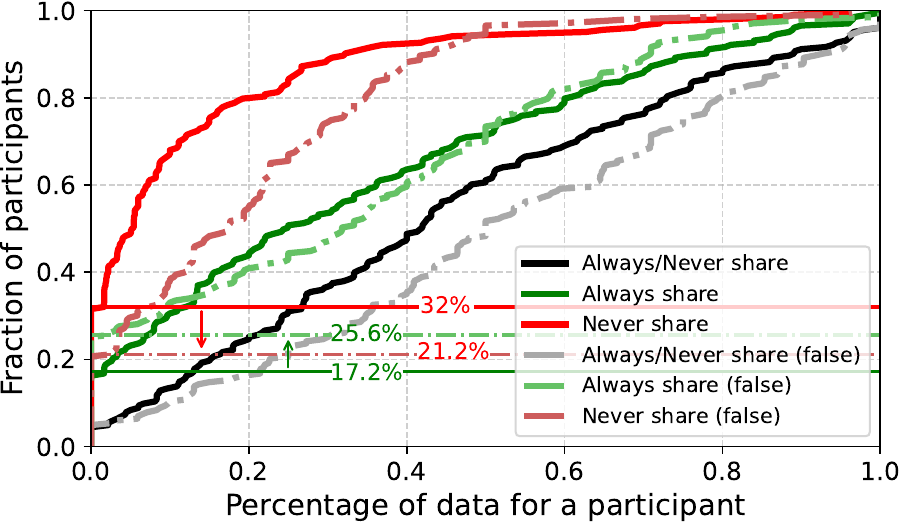}
    \caption{Distribution of participant permission preferences. The \textit{false} label indicates data sharing instances where unnecessary data types were presented to participants (along with the necessary data types).}
    \label{fig:cdf_allow_disallow}
\end{figure}

As we explore automating users' data sharing permissions decisions, we first investigate permission preferences that participants expressed to the AI agent. 
We consider \textit{yes, always share} and \textit{no, never share} permissions, as permissions expressed for the AI agent to learn participant preferences. 
Figure~\ref{fig:cdf_allow_disallow} presents a distribution of participants and their permission preferences. 
We note that 95.1\% of the participants express \textit{always share} permission decisions to AI agents for one or more data sharing decisions. %
The proportion of permission decisions is higher for sharing data than for not sharing data, which suggests that users may engage in over-permissioning (discussed in Section~\ref{subsubsection:permission-practices}).
Specifically, 82.8\% and 68.0\% of participants let AI agents to automatically share and not share data at least once, respectively. 

Since real-world AI agents can make mistakes and over-collect data, we next explore how user permission preferences change when AI agents make mistakes. 
Dotted lines in Figure~\ref{fig:cdf_allow_disallow} present distributions of participants and their permission preferences when the AI agent makes mistakes. 
We find that participants tend to express more permission decisions for not sharing data (i.e., \textit{never share}). %
Specifically, the number of participants who never express \textit{never share} permission decision decreases from 32\% (when participants are presented with necessary data) to 21.2\% (when participants are presented with unnecessary data, along with necessary data). %
This suggests that AI agent mistakes prompt users to assess their permission decisions, and many users are able to identify unnecessary data permissions, and convey to the AI agent to never share that information.

We note that the number of participants granting permissions to automatically share data decreases. 
Specifically, the number of participants who never express \textit{always share} permissions increases from 17.2\% (when participants are presented with necessary data) to 25.6\% (when participants are presented with unnecessary data, along with necessary data).
When participants do express automatic sharing/non-sharing permissions, we do not observe substantial changes in their \textit{always share} preferences.
We surmise that when some users observe agent mistakes, they are more cautious in expressing preferences for automatic data sharing.

\subsubsection{We observe that over-permissioning is substantially more common than under-permissioning. We also note that while participants struggle with providing appropriate permissions and often over-share, they are mostly cautious about sharing sensitive data}
\label{subsubsection:permission-practices}
Under- and over-permissioning have been persistent problems in prior systems, such as mobile platforms, where users often granted too many or too few permissions due to poor understanding, misleading interfaces, or decision fatigue~\cite{felt2012android,kelley2012conundrum}, leading to privacy risks and/or broken functionality~\cite{almuhimedi2015your,wijesekera2015android}.
Motivated by these challenges, we next examine whether similar issues arise in the context of AI agents, where the opacity of AI agents' execution may make permission alignment even more challenging.

\begin{figure}[t]
    \centering
    \includegraphics[width=0.4\textwidth]{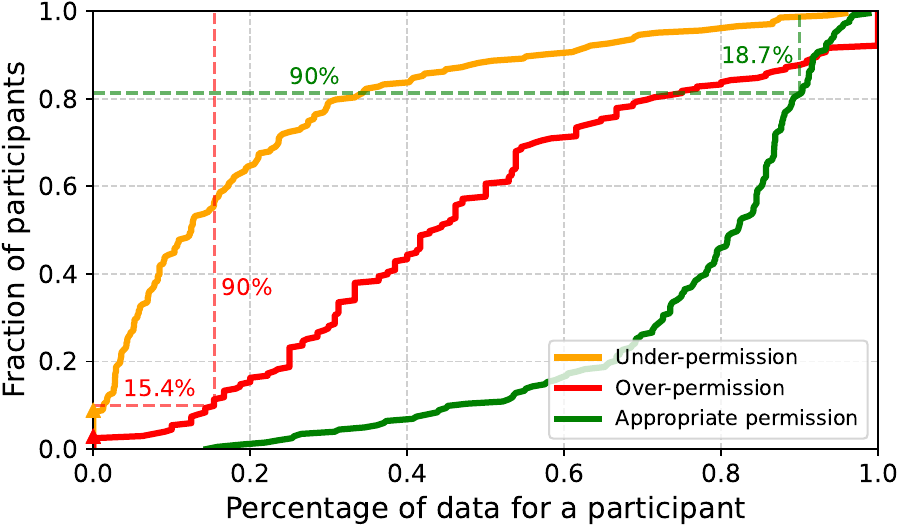}
    \caption{Distribution of \textit{under-}, \textit{over-}, and \textit{appropriate} data sharing permission rate of each participant.}
    \label{fig:cdf_over_under_permission}
\end{figure}

Figure~\ref{fig:cdf_over_under_permission} presents the distribution of under-, over-, and appropriate data sharing permission rates of each user.\footnote{\textit{Appropriate permission} considers participant that provided only the data that was necessary (as determined during query curation$\S$~\ref{subsubsection:query-curation}) for solving the query, \textit{over-permission} considers participants that shared data that was not required to resolve the query, and \textit{under-permission} considers participant that withheld data that was necessary to address the query.}
We find that only 8.9\% of participants never engage in under-permissioning and 3.0\% of participants never engage in over-permissioning. 
We observe that over-permissioning is substantially more common than under-permissioning, where 90\% of participants engaging in over-permissioning for 15.4\% or more data sharing requests they encounter.
None of the participants give appropriate permissions for all of the data types on which they were prompted, and only 18.7\% of the users give appropriate permissions for 90\% or more data types on which they were prompted.

Through more in-depth analysis, we find that participants' data sharing behavior is impacted by the sensitivity of data. For example, highly sensitive information such as \textit{SSN} exhibits an under-permission ratio of about 46.3\%, meaning participants are cautious with sharing such data.
In contrast, for relatively less sensitive data, such as \textit{Meeting Details}, which has an appropriate permission of 98.4\%, participants are much more comfortable granting access.
For ambiguous but non-sensitive data types, such as \textit{Workspace Name} (generally used by Slack), participants tend to over-permission for 88.9\% of the instances, likely because they may be uncertain about the actual need of this information.

Additionally, we analyze some of the most frequently under-permissioned data types and find that participants are less likely to share sensitive personal information.
For example, \textit{Child Name} was not shared 65.1\% of the time, \textit{Work Information} by 57.4\%, and \textit{Passport Information} by 54.0\% across all data permission requests.
These data types are often essential for tasks, such as identity verification or financial transactions.
Importantly, participants who did not share these data types frequently reported strong privacy concerns in relevant domains.
For instance, 50.0\% (15) of participants who did not share \textit{Account Password} and 36.8\% (25) who did not share \textit{SSN} listed \textit{Finance} as one of their privacy concerning domains.
This suggests that participants are making intentional privacy-conscious choices, particularly in areas they value most.
While some of these data types, such as \textit{SSN} or \textit{Passport Information}, are often essential for providing functionality, we find that others, such as \textit{Child Name} and \textit{Travel Details}, could be reasonably substituted with dummy values in some scenarios to maintain functionalities while respecting users' privacy.

\begin{table}[t]
\centering
\footnotesize
\begin{tabular}{l|cc|cc}
\toprule
              & \multicolumn{2}{c|}{\textbf{All}} & \multicolumn{2}{c}{\textbf{Concerning domain}} \\ \midrule
\textbf{Domain}        & \textbf{Always}  & \textbf{Never}           & \textbf{Always}        & \textbf{Never}               \\ \midrule
\textbf{Entertainment}    & 55.6\%  &2.9\%            & 60.5\%     & 2.6\%     \\
\textbf{Health\&Fitn.} & 48.9\%   & 7.7\%            & 44.6\% & 9.7\%              \\
\textbf{Smart Home}    & 41.6\%    & 9.3\%            & 43.5\%  & 7.6\%               \\
\textbf{Travel}        & 36.0\%      & 10.5\%           & 20.3\% & 3.8\%     \\
\textbf{Shopping}      & 32.2\%     & 12.0\%           & 24.0\%  & 9.3\%              \\
\textbf{Work\&Produc.} & 30.8\%     & 8.6\%            & 27.6\% & 7.9\%               \\
\textbf{Social}        & 30.4\%     & 8.1\%            & 33.6\%  & 9.9\%               \\
\textbf{Finance}       & 22.2\%  & 16.9\%           & 22.4\%       & 13.9\%     \\ 
\midrule
\textbf{All}       & 33.8\%     & 10.7\%           & 29.2\%     & 10.8\%     \\
\bottomrule
\end{tabular}
\caption{Permission preferences (avg.) for always and never share permissions. \textit{Concerning domain} columns contain data only from participants who mark the corresponding domain as concerning.}
\label{tab:data_sharing_preferences_domain}
\end{table}

\subsubsection{Participants' permission preferences vary even within various contexts/domains, thus requiring finer contextual considerations. Participants are also able to identify and correct AI agents' mistakes in sharing data}
\label{section:contextual_analysis}
To make automatic decisions on users' behalf, an AI agent needs to understand the factors that influence users' permission decisions.
We dive into several factors. %

\vspace{1mm}\noindent\textbf{Communication Context/Domain.} Significant prior research has identified the ``context'' of the communication as a key factor in user decision-making around data sharing~\cite{bagdasarian2024airgapagent,abdi2021privacy,barth2006privacy,jia2017contexlot,nissenbaum2004privacy}. %
Table~\ref{tab:data_sharing_preferences_domain} presents the eight contexts (which we refer to as domains) for which we record user preferences. 
We observe considerable variation in participants' permission preferences in \textit{always share} and \textit{never share} permissions across different domains.
For data sharing, we note that participants express the most \textit{always share} permissions (i.e., 55.6\%) for the \textit{Entertainment} domain and the least permissions (i.e., 22.2\%) for the \textit{Finance} domain.
For the \textit{never share} permissions, the inverse is true, i.e., participants express their preferences to not share data for only 2.9\% of cases within the \textit{Entertainment} domain, but in 16.9\% of cases with the \textit{Finance} domain.
We also observe that participants tend to give fewer \textit{always share} permissions, for domains that they self-report as privacy concerning.
This variation is particularly noticeable for \textit{Travel} and \textit{Shopping} domains, where \textit{always share} decisions drop by 15.7\% and 8.2\%, respectively.

\begin{table}[t]
\resizebox{\columnwidth}{!}{
\centering
\begin{tabular}{l|l|cccc}
\toprule
\textbf{Domain} & \textbf{Tool} & \textbf{Always} & \textbf{Yes (Once)} & \textbf{No (Once)} & \textbf{Never} \\
\midrule
\multirow{3}{*}{Entertainment}
& Movie Database & \gheatbg{58.3} & \gheatbg{33.7} & \rheatbg{5.9} & \rheatbg{2.1} \\
& Web & \gheatbg{34.4} & \gheatbg{52.5} & \rheatbg{4.9} & \rheatbg{8.2} \\
& Apple Music & \gheatbg{68.3} & \gheatbg{30.2} & \rheatbg{1.6} & \rheatbg{0.0} \\
\midrule
\multirow{5}{*}{Travel}
& Weather & \gheatbg{50.8} & \gheatbg{37.9} & \rheatbg{5.1} & \rheatbg{6.1} \\
& Local Search & \gheatbg{42.9} & \gheatbg{39.0} & \rheatbg{9.9} & \rheatbg{8.2} \\
& Travel Booking & \gheatbg{32.6} & \gheatbg{45.4} & \rheatbg{9.6} & \rheatbg{12.5} \\
& Calendar & \gheatbg{43.9} & \gheatbg{43.9} & \rheatbg{5.3} & \rheatbg{6.9} \\
& Cloud Drive & \gheatbg{17.8} & \gheatbg{56.3} & \rheatbg{9.6} & \rheatbg{16.4} \\
& Email & \gheatbg{34.4} & \gheatbg{50.8} & \rheatbg{9.0} & \rheatbg{5.7} \\
\midrule
\multirow{4}{*}{Work\&Produc.}
& Slack & \gheatbg{34.9} & \gheatbg{47.6} & \rheatbg{7.9} & \rheatbg{9.5} \\
& Email & \gheatbg{27.7} & \gheatbg{53.6} & \rheatbg{8.3} & \rheatbg{10.4} \\
& Calendar & \gheatbg{43.8} & \gheatbg{47.4} & \rheatbg{5.6} & \rheatbg{3.2} \\
& Cloud Drive & \gheatbg{22.6} & \gheatbg{64.5} & \rheatbg{7.3} & \rheatbg{5.6} \\
\midrule
\multirow{5}{*}{Health\&Fitn.}
& Fitness Tracking & \gheatbg{48.7} & \gheatbg{35.4} & \rheatbg{9.0} & \rheatbg{6.9} \\
& PregnancyPal & \gheatbg{41.0} & \gheatbg{42.6} & \rheatbg{6.0} & \rheatbg{10.4} \\
& Nutrition\&Diet & \gheatbg{46.2} & \gheatbg{37.3} & \rheatbg{7.5} & \rheatbg{9.0} \\
& Hospital Booking & \gheatbg{54.0} & \gheatbg{33.9} & \rheatbg{6.5} & \rheatbg{5.6} \\
& Calendar & \gheatbg{54.8} & \gheatbg{35.5} & \rheatbg{4.8} & \rheatbg{4.8} \\
\midrule
\multirow{4}{*}{Social}
& SMS & \gheatbg{29.2} & \gheatbg{51.4} & \rheatbg{7.6} & \rheatbg{11.7} \\
& Cloud Drive & \gheatbg{27.9} & \gheatbg{54.1} & \rheatbg{13.1} & \rheatbg{4.9} \\
& Instagram & \gheatbg{26.2} & \gheatbg{52.5} & \rheatbg{14.8} & \rheatbg{6.6} \\
& Tinder & \gheatbg{32.2} & \gheatbg{50.0} & \rheatbg{10.0} & \rheatbg{7.7} \\
\midrule
\multirow{2}{*}{Finance}
& Banking & \gheatbg{24.8} & \gheatbg{50.6} & \rheatbg{11.3} & \rheatbg{13.2} \\
& Tax Management & \gheatbg{22.6} & \gheatbg{50.2} & \rheatbg{9.2} & \rheatbg{18.0} \\
\midrule
\multirow{1}{*}{Shopping}
& Amazon & \gheatbg{32.2} & \gheatbg{45.0} & \rheatbg{10.7} & \rheatbg{12.0} \\
\midrule
\multirow{1}{*}{Smart Home}
& Lighting Control & \gheatbg{54.8} & \gheatbg{32.3} & \rheatbg{3.2} & \rheatbg{9.7} \\
\bottomrule
\end{tabular}
}
\caption{Permission preferences across domains and tools.}
\label{tab:tool_analysis}
\end{table}

\vspace{1mm}\noindent\textbf{Differences Within Communication Contexts/Domains.}
Next, we analyze the variance in participants' permission preferences within domains. %
We present the breakdown of all participants' permission preferences for data-sharing for different tools within different domains in Table~\ref{tab:tool_analysis}.
Overall, we note that participant preferences on data sharing for tools have differences within domains.
Barring domains with single tools, \textit{Entertainment} and \textit{Travel} have the highest standard deviation of 14.2\% and 10.5\% for always sharing, and \textit{Finance} has the lowest standard deviation of 1.1\%.
In the case of \textit{Entertainment}, \textit{Web} browsing tool gets the least \textit{always share} permission preferences, and in the case of \textit{Travel}, \textit{Cloud Drive} gets the least \textit{always share} permission preferences.
Our investigation reveals that, in the context of \textit{Travel}, \textit{Cloud Drive} is always used to access users' travel documents (e.g., passport scans, visa records).
While participants allow access to these documents as they are essential for some travel tasks, they do not prefer AI agents to automatically retrieve and share such sensitive data.

For one-time sharing, \textit{Entertainment} and \textit{Finance} have the highest and lowest standard deviation of 9.8\% and 0.2\%, respectively.
For never sharing and one-time non sharing permission preferences, the standard deviation between tools across domains remains mostly stable, and does not exceed 3.9\% for never sharing and 2.8\% for one-time non sharing.

These results indicate that permission preferences can vary even within domains, and thus granular context consideration is necessary for permission inferences, i.e., considering coarse context categories for understanding permission preferences across users may be insufficient. %
Our permission prediction model (in Section~\ref{subsubsection:hybrid-model}) thus encodes contextual information at a finer granularity.

\begin{table}[t]
\setlength{\tabcolsep}{9pt}
\resizebox{\columnwidth}{!}{
\centering
\begin{tabular}{l|l|cc|cc}
\toprule
\textbf{GT} & \textbf{PA} & \textbf{Always} & \textbf{Yes (Once)} & \textbf{No (Once)} & \textbf{Never} \\
\midrule
Necessary              & Necessary             & 41.8\%                    & 47.1\%                         & 4.9\%                         & 6.1\%                     \\
Necessary              & Unnecessary            & 31.1\%                     & 36.0\%                         & 15.2\%                        & 17.7\%                    \\
Unnecessary           & Necessary             & 30.2\%                     & 48.2\%                         & 12.0\%                        & 9.5\%                     \\
Unnecessary            &Unnecessary            & 16.6\%                     & 21.0\%                         & 21.1\%                        & 41.4\%                 \\
\bottomrule
\end{tabular}
}
\caption{Participant permission prefs. for perceived necessary/unnecessary data (PA), along with ground truth (GT).}
\label{tab:necessary_align}
\end{table}

\vspace{1mm}\noindent\textbf{Differences When Users Align and Misalign With the AI Agent.}
As we ask participants in our user study to select the \textit{necessary data} that the AI agent will need to address a query, we develop a notion of participants developing an \textit{understanding} of the AI agents' execution. %
This setup allows us to analyze how user permission preferences may vary when they have some understanding of the AI agent's execution. 
Table~\ref{tab:necessary_align} presents participants' permission preferences for perceived necessary and unnecessary data, along with the ground truth (as determined in Section~\ref{subsubsection:query-curation}).
We observe that when participant perceptions of necessary and unnecessary data align with the ground truth (1st and 4th row in Table~\ref{tab:necessary_align}), the rate of appropriate permission decisions (allowing necessary data and denying unnecessary data) is high.
In such cases, participants are more likely to express appropriate permissions, with an 88.9\% and 62.5\% correct decision rate for sharing and not sharing.
We also note that, even when participants believe certain necessary data is unnecessary (2nd row in Table~\ref{tab:necessary_align}), they are still able to provide appropriate permissions, likely because data selection for specific queries provides more context for making an informed decision. 
Conversely, when the agent makes incorrect decisions, i.e., presents unnecessary data as necessary (3rd row in Table~\ref{tab:necessary_align}), 78.2\% of times users share data with the AI agent, which implies that many users tend to believe the AI agent even when they are incorrect.

\subsubsection{Participant demographics information (e.g., age) and self-reported metrics (e.g., privacy consciousness) correlate with their permission decision preferences}
\label{subsec:factor_analysis}
Next, we analyze participants' demographic information (i.e., age, education, and sex) and self-reported metrics (i.e., AI familiarity, AI usage frequency, AI trust, and privacy consciousness) to analyze how these factors influence participants' permission preferences.
As for demographics, we observe that the tendency to give \textit{Always Allow} data sharing permissions decreases with participant age, correlating with generational differences in privacy attitudes as also observed by prior work~\cite{abdi2021privacy}. 
We observe that there is no substantial difference in the permission preferences of male and female participants. 
As for self-reported metrics, higher AI familiarity and usage correlate with increased data-sharing preferences. 
We provide a more detailed analysis of demographics and self-reported metrics in Appendix~\ref{sec:ana_demo}.

\begin{figure}[t]
    \centering
    \includegraphics[width=0.45\textwidth]{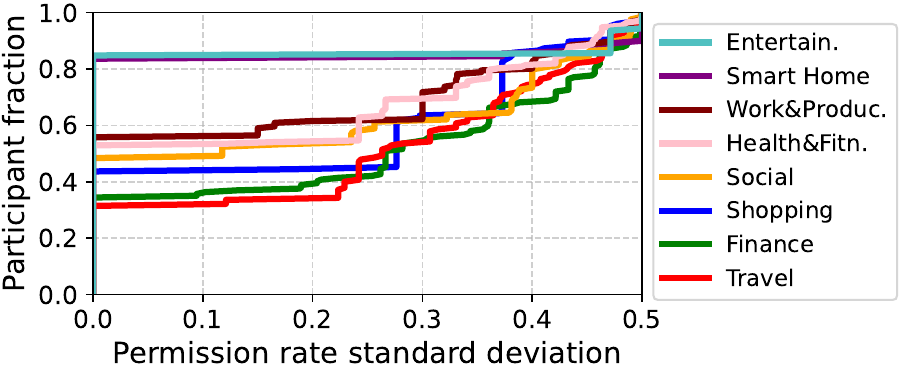}
    \caption{Distribution of standard deviation (SD) in participant permission preferences within domains. SD of 0 implies that participants choose the same preference within a domain, 0.1 implies there are 2 or 3 data types that do not align with other data types within a domain.}
    \label{fig:variance_domain}
\end{figure}

\subsubsection{Individual participants' permission preferences within a domain are often consistent; however, for some privacy-conscious participants, there is a high variance in their permission decisions}
\label{subsec:learn_preferences-intra-domain}
For an AI assistant to learn users' permission preferences and apply them in various situations, an important criterion is that users' decision-making is predictable and consistent.
To that end, we analyze the consistency in participants' permission decisions. 
From Figure~\ref{fig:variance_domain}, we note that within a domain, participants' permission decisions are often consistent. 
The standard deviation is lowest for the \textit{Entertainment} and \textit{Smart Home} domains and highest for the \textit{Travel}, \textit{Finance}, and \textit{Shopping} domains. 
Notably, many participants have entirely consistent data-sharing preferences within specific domains.
There are 84.5\%-85.6\% of participants who are willing to either allow or deny all data permission for the \textit{Entertainment} and \textit{Smart Home} domains, 49.2\%-56.4\% for the \textit{Work \& Productivity}, \textit{Health \& Fitness}, and \textit{Social} Domains, and 32.1\%-44.4\% for the \textit{Shopping}, \textit{Finance}, and \textit{Travel} domains.

Additionally, we notice that some participants have relatively high intra-domain preference variance.
We therefore analyze the top 10 participants with high intra-domain preference standard deviation and find that the wide variation for these participants can be traced to several key personal attributes, particularly AI trust levels, privacy concerns, and AI familiarity. 
Specifically, 8 participants report low to moderate trust in AI, even though 7 participants use AI frequently, and all rate privacy as very important.
For these participants, the sensitivity of data types seems to matter when they make permission decisions.
For example, within the \textit{Finance} domain, participants often deny access to highly sensitive personal data such as SSNs, bank account numbers, and online account credentials, while allowing access to relatively less sensitive data, such as tax filing status, savings goals, or payment notes.
Such a distinction leads to varied preferences for participants even within a single domain.

\begin{figure}[t]
    \centering
    \includegraphics[width=0.45\textwidth]{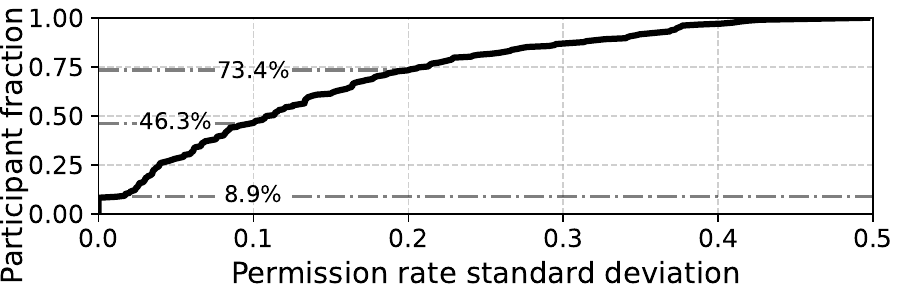}
    \caption{Distribution of SD in participant permission preferences across domains. SD of 0.1 means that preferences are mostly consistent, and SD of 0.2 means that only 1 or 2 domains have a noticeable difference with others.}
    \label{fig:cdf_preference__standard_deviation}
\end{figure}

\subsubsection{Participants' permission preferences tend to follow similar patterns, presenting an opportunity to predict individual user preferences by leveraging insights from other users}
\label{subsec:learn_preferences-inter-domain}
In a real-world deployment, we anticipate that users will encounter a range of scenarios that may not arise during the training of a personal permission assistant. 
For example, to not fatigue the users, a deployed permission assistant may record a user's preferences in one, two, or a handful of domains, but users' day to day questions may span more domains.
Thus, we seek to understand how transferable users' permission decision making patterns are across contexts. 
Figure~\ref{fig:cdf_preference__standard_deviation} presents the standard deviation of permission allowance rate across domains for users (Figure~\ref{fig:heatmap_stacked} in Appendix~\ref{sec:heatmap} shows the distribution of average permission allowance rates across domains for users.).
We note that 46.3\% of participants have a standard deviation below 0.1, which means their permission is mostly consistent for all the tested domains; 73.4\% of participants have a standard deviation below 0.2, which means only one or two domains have a noticeable difference with other domains.
Moreover, we find that 8.9\% of participants maintain consistent permission preferences across all tested domains.
We also find that over 35.5\% of participants (not explicitly represented in Figure~\ref{fig:cdf_preference__standard_deviation}) show zero preference variance in at least half of the domains they interact with, suggesting that they tend to grant similar permissions across different contexts.

\begin{figure}[t]
    \centering
    \includegraphics[width=0.45\textwidth]{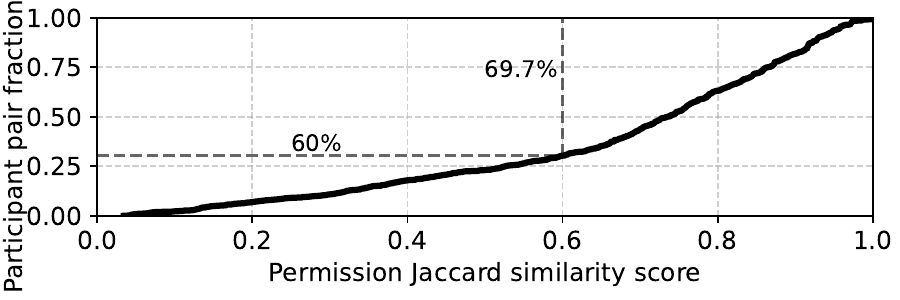}
    \caption{Jaccard similarity for participant pairs within groups. We group participants by the query set they answer.}
    \label{fig:jaccard_similarity}
\end{figure}

We also explore the consistency in permission preferences for data types across participants, as it can allow us to learn preferences from groups of users with similar preferences and apply them to other users. %
Figure~\ref{fig:jaccard_similarity} presents the Jaccard similarity between participants' permission preferences on the same data permission requests.
We note that participants' permission preferences for the same permission requests are not unique and in fact, resemble many other users. 
For example, 69.7\% of participant pairs (two participants answering the same permission requests) have similar data sharing preferences for 60\% or more data types. 
To further understand user variance and consistency in data permission preferences, we analyze which data types exhibit the highest and lowest variability across participants.
As shown in Table~\ref{tab:10_datatype_std}, high-variance types like \textit{Account Credentials}, \textit{Employment Details}, and \textit{Driver License Number} (with standard deviations of 0.49) indicate strong disagreement---likely due to the sensitivity of data.
In contrast, low-variance types such as \textit{Music Listening History} and \textit{Hobbies and Interests} (with standard deviations as low as 0.125) reflect broad agreement, possibly because they are seen as low-risk and frequently shared in everyday contexts.
These patterns suggest that preferences for some data types are relatively predictable, while others show diverse responses and are more challenging to predict.

\begin{table}[t]
\resizebox{\columnwidth}{!}{
\centering
\begin{tabular}{lc|lc}
\toprule
\multicolumn{2}{c|}{Top 10 high std. dev. data types}     & \multicolumn{2}{c}{Top 10 low std. dev. data types}    \\
\midrule
Data type              &  Std. dev.    & Data type                &  Std. dev.    \\
\midrule
Account Credentials    & 0.498 & Music Listening History  & 0.125 \\
Employment Details     & 0.498 & Hobbies and Interests    & 0.178 \\
Driver License Number  & 0.489 & Relationshop Preferences & 0.212 \\
Travel Itinerary       & 0.486 & Test/Diagnostic Results  & 0.244 \\
Passport Documents     & 0.485 & Fitness Goal             & 0.244 \\
Investment Information & 0.483 & Movie Preferences        & 0.246 \\
Payment Method Details & 0.482 & Personal Biography       & 0.248 \\
Bank Account Details   & 0.478 & Travel Preferences       & 0.251 \\
Social Security Number & 0.475 & Accommodation Details    & 0.255 \\
Sender Email Address   & 0.461 & Travel Destination       & 0.259 \\
\bottomrule
\end{tabular}}
\caption{Data types with high and low standard deviation in participants' data permission preferences.}
\label{tab:10_datatype_std}
\end{table}

%% file: 5_predicting_preferences.tex
\section{Predicting User Permission Preferences}
\label{sec:model}

In this section, we explore the feasibility of predicting users' permission decisions by leveraging data from our user study. 
We explore the potential of using \textit{individual} user data as well as leveraging data from \textit{other users} in developing an accurate permission prediction model. 
We also explore the role of various factors, such as the domains/contexts, individual data types, and variance in user permission preferences, in the accuracy of permission prediction.

We anticipate a sustained progression in permission modeling for AI agents over the next several years, during which a wide variety of models will be explored.
In this paper, we focus more on the implications from our user study results and present our \textit{permission assistant} as a proof of concept and/or initial exploration of the design space, rather than a definitive solution.

\subsection{Designing a Permission Prediction Model}
\label{subsection:prediction}
Our findings in Section~\ref{sec:preferences_findings} indicate that although various factors influence participants' permission decisions, these influences ultimately result in relatively consistent participant permission decisions.
Thus, we investigate whether users' prior permission decision history, their demographics, and other self-reported attributes can be used to predict their future permissions decisions.

A key challenge is that we possess limited data on users, which may make it challenging to train a classifier that attains a high accuracy.
To this end, we explore a hybrid machine learning framework that relies on LLM-based in-context learning~\cite{nori2023can} and collaborative filtering~\cite{he2020lightgcn}.
We rely on LLM-based in-context learning because it can attain a high accuracy even with a handful of examples (i.e., ``few-shot'')~\cite{nori2023can,dong2024survey}.
We rely on collaborative filtering because it allows us to learn from other users with similar permission decision history~\cite{he2017neural,he2020lightgcn}, thus complementing the limited permission history we possess on individual users.

\subsubsection{In-context learning} 
\label{subsec:in-context-learning}
As we observe in Section~\ref{subsec:learn_preferences-intra-domain} and~\ref{subsec:learn_preferences-inter-domain}, participants' permission decisions remain consistent within a domain and are transferable across domains, we first attempt to learn preferences only from users' own permission decisions. 
To that end, we design an LLM-based in-context learning framework using OpenAI's o3-mini reasoning-based LLM. %
To condition the LLM, we rely on role-based prompting, user demographics, other self-reported information, and user permission history.

Since our goal is to design an assistant that will assist users in their permission decision making, we adopt \textit{permission assistant} as a role for the LLM.
As for the demographics we include, users' age range, education, and gender. %
For self-reported information, we mainly consider users' AI familiarity (e.g., usage frequency, usage purposes) and privacy consciousness (e.g., value to privacy, trust in AI). %
User permission history includes the query, data type, and the name of the tool or AI agent requesting the data, along with the user's permission decision.
As LLMs take natural language input, we encode these features as natural language instructions.
For example, demographic information is encoded as follows: \textit{a male in the 45–54 age group with a bachelor's degree}.
As an output, we condition the LLM to provide a prediction label along with a confidence score, which prior research has shown to be reliable~\cite{pawitan2025confidence}.

\subsubsection{Collaborative filtering} 
Motivated by our observation in Section~\ref{subsec:learn_preferences-inter-domain}, where groups of users exhibit similar permission-granting behaviors across domains and scenarios, we explore collaborative filtering in predicting user preferences. %
Collaborative filtering is widely adopted in recommendation systems due to its effectiveness in capturing implicit user-user similarities based on commonalities in prior user preferences~\cite{he2017neural,he2020lightgcn}.
We model user preferences as an adjacency matrix, where columns represent data types across contexts, rows represent users, and values in each cell represent either positive (sharing) or negative (non-sharing) permission preference. 
Since users do not provide their preferences on all data types, several of the cell values are empty, and our classification task is to predict the values of the cells.
We rely on model-based collaborative filtering, which uses a light graph convolution network (LightGCN)~\cite{he2020lightgcn}.

LightGCN constructs a graph where users and permission requests (with context) are nodes, and edges represent observed permission preferences.
It learns embeddings for users and permission requests by iteratively averaging information from their connected neighbors.
The final prediction score is computed using the dot product between a user and a permission request embedding, indicating the likelihood that the user would grant the permission.
This score is then converted into a predicted label (i.e., allow or deny) by applying a threshold.
The prediction threshold is configurable for collaborative filtering, which we currently set to equalize FPR and FNR (Appendix~\ref{sec:cf_raw_metrics} provides more details).

\subsubsection{Hybrid model using in-context learning and collaborative filtering}
\label{subsubsection:hybrid-model}
We next explore combining the in-context learning and collaborative filtering models to incorporate learning from both: (1) individual user preferences, which in-context learning incorporates, and (2) similar user preferences, which collaborative filtering incorporates.
We combine these models by extending our in-context learning framework (described in Section~\ref{subsec:in-context-learning}).
Specifically, we include the results from a collaborative filtering model as textual descriptions for the LLM.
For example, a recommended permission decision is represented as follows: \textit{$<$Query: Can you retrieve my tax filing details from last year?; Tool: Tax Management; Data Type: SSN; Decision: Deny$>$}.
Since preferences for different participants can vary within domains (as shown in Section~\ref{section:contextual_analysis}), such representation at the query level (i.e., finer granularity) allows us to naturally capture consistent preferences for individual participants. %

While integrating collaborative filtering results, we only consider high-confidence permission predictions. %
We consider high-confidence predictions to be the ones where both the false positive rate (FPR) and false negative rate (FNR) are below 5\%, which covers 35.0\% of the permission requests.
Note that in our testing, we tried including all CF predictions (i.e., both high- and low-confidence predictions) in our hybrid model, but that deteriorates model accuracy.

Note that our collaborative filtering alone is incapable of making predictions on previously unseen data, and our in-context learning model only used data from individual users. 
Our hybrid model allows collaborative filtering and in-context learning to complement each other.
As a result, it yields a personalized predictor for each user (i.e., as many personalized instances as users), composed of shared user-specific CF parameters as well as an in-context learning model trained on individual user data.

\begin{figure*}[t]
    \centering
    \begin{minipage}[t]{0.74\linewidth} %
        \centering
        \begin{subfigure}[t]{0.32\linewidth}
            \centering
            \includegraphics[height=3.3cm]{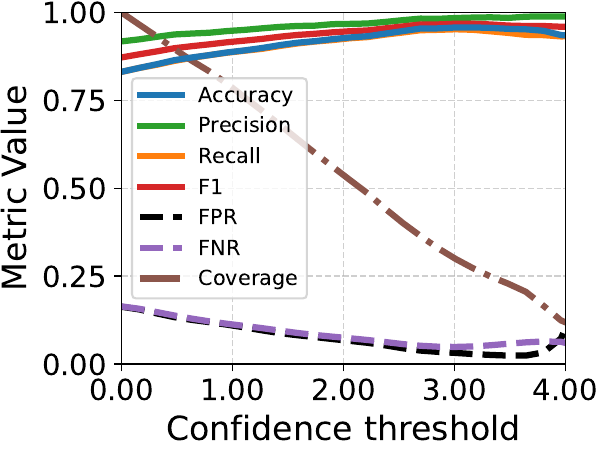}
            \caption{CF}
            \label{fig:cf_metrics_vs_threshold}
        \end{subfigure}
        \hfill
        \begin{subfigure}[t]{0.32\linewidth}
            \centering
            \includegraphics[height=3.3cm]{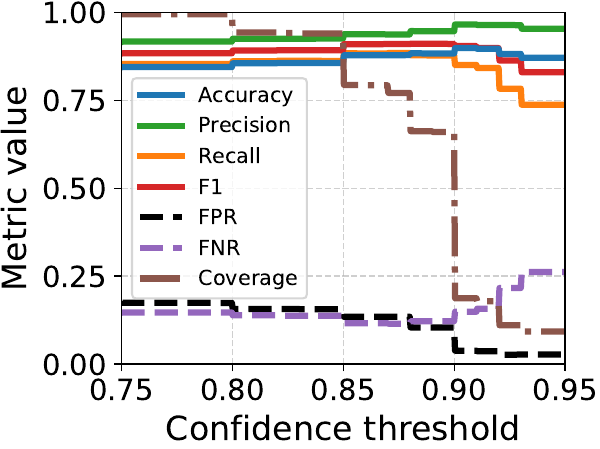}
            \caption{IC}
            \label{fig:llm_metrics_vs_threshold}
        \end{subfigure}
        \hfill
        \begin{subfigure}[t]{0.32\linewidth}
            \centering
            \includegraphics[height=3.3cm]{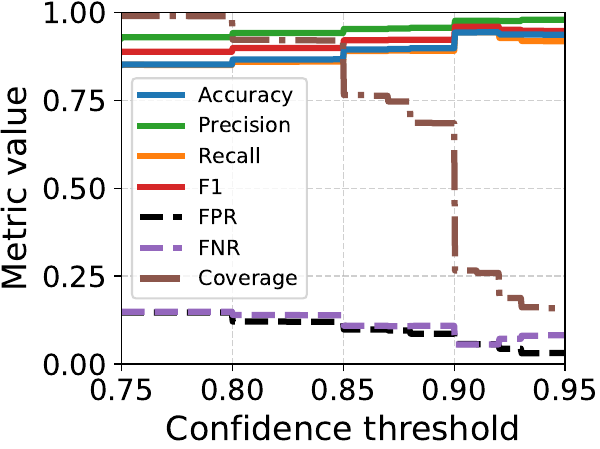}
            \caption{IC \& CF hybrid}
            \label{fig:metrics_vs_threshold}
        \end{subfigure}
        \caption{Distribution of classification metrics. For IC \& hybrid models, the confidence score is reported by the LLM. For CF, the confidence score is the difference b/w prediction thresholds that reduce FPR and FNR, with larger values implying higher confidence.}
        \label{fig:combined_metrics}
    \end{minipage}
    \hfill
    \begin{minipage}[t]{0.24\linewidth}
        \centering
        \includegraphics[height=3.3cm]{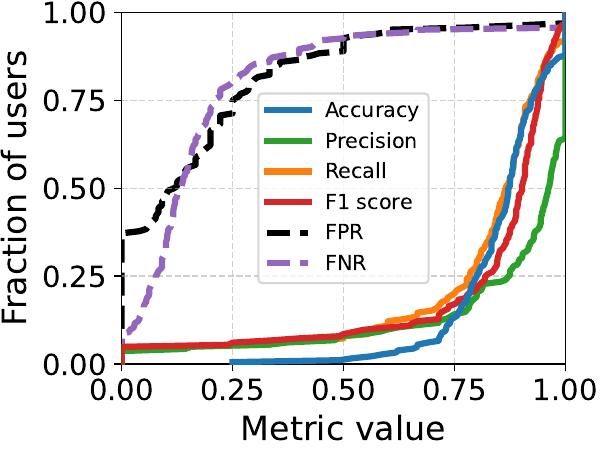} \vspace{1.8mm}
        \caption{Distribution of hybrid model metrics over the fraction of users.}
        \label{fig:cdf_all_metrics}
    \end{minipage}
    \vspace{-1.8mm}
\end{figure*}

\subsection{Evaluation}
\label{sec:evaluation}
\subsubsection{Datasets and metrics}
We construct our training and testing datasets using responses collected from the user study to evaluate the effectiveness of our permission prediction model. 
We only consider decisions marked as \textit{yes, always share} or \textit{no, never share} as ground truth for sharing and non-sharing data. %
Additionally, we exclude participants who specified always/never sharing preferences for fewer than five scenarios/queries.
After filtering, our dataset contains 7{,}563 permission decisions from 181 participants, including 5{,}244 allowance and 2{,}319 denial responses.
We define a positive outcome as permission being granted and a negative outcome as permission being denied. 
For model training and evaluation, we perform 5-fold cross-validation. %
Specifically, the user-answered questions are split into five folds; in each iteration, we train on four folds and test on the remaining fold. This is repeated five times, so every question is tested exactly once and is never used for training and testing in the same iteration. This setup allows us to evaluate the models on unseen data.

\subsubsection{Classification confidence threshold}
We start by comparing model configurations: (1) a model trained on individual user data using in-context (IC) learning, (2) a collaborative filtering (CF) model trained on the entire permission-granting history of all users, and (3) a hybrid of in-context learning and collaborative filtering. 
Figure~\ref{fig:combined_metrics} presents the distribution of classification metrics for various model configurations. 
Overall, we attain the highest accuracy (95.8\%) for CF with a confidence threshold of 3.17. 
However, we compromise on \textit{coverage}\footnote{We define \textit{coverage} as the fraction of permission requests for which the model can recommend an action with confidence above a specified threshold. Lower coverage, however, also means that users must make more decisions manually, and only high-confidence cases are automated.} as we are only able to make predictions on 27.3\% of the data. 
For IC, we attain an accuracy of 89.9\% with a confidence threshold of 0.90, and a coverage of only 18.8\%.
For the hybrid model of IC and CF, we attain an accuracy of 94.4\% with a confidence threshold of 0.91, and a coverage of 25.9\%.
These trends indicate that, as the confidence threshold gets stricter, it results in precise predictions but reduces the coverage. %

\begin{table}[t]
\setlength{\tabcolsep}{12pt}
\begin{tabular}{lccc}
\toprule
Metric          & CF (\%)  &  IC (\%)         &  IC \& CF (\%)       \\ \midrule
Accuracy   & 83.3±1.4 & 84.4±0.7          & \textbf{85.1±0.4} \\
Precision & 91.9±0.5 & 91.5±1.1          & \textbf{92.8±0.7} \\
Recall     & 83.3±1.4 & \textbf{85.4±1.5} & 85.2±1.2 \\
F1 score  & 87.4±1.0 & 88.3±0.7          & \textbf{88.8±0.5} \\
FPR        & 16.7±1.5 & 17.9±2.3          & \textbf{15.2±2.1} \\
FNR        & 16.7±1.4 & \textbf{14.6±1.5} & 14.8±1.2 \\ \bottomrule
\end{tabular}
\caption{Classification metrics with full coverage across different model configurations.\label{tab:model_compare}}
\end{table}

\vspace{1mm}\noindent\textbf{Classification Accuracy.}
\label{paragraph:classification-accuracy}
To make classification decisions for all data, i.e., for 100\% coverage, across all model configurations, we are able to achieve an accuracy of 83.3\%, 84.4\%, and 85.1\%, for CF, IC, and the hybrid IC and CF model.
Table~\ref{tab:model_compare} presents other classification metrics for full data coverage.
We note that the hybrid model outperforms individual model configurations on all metrics, except for recall and FNR.
Notably, the FPR decreases by 2.7\% when recommendations from the CF model are taken into account. %
These results indicate that incorporating contextual examples from the CF model enhances prediction accuracy.
However, as we note in Section~\ref{subsubsection:hybrid-model} that these gains strongly correlate with the quality of CF predictions.

As permission assistants will make predictions to share data on behalf of users, a high precision and lower FPR may be desired, which requires a compromise on model coverage.
Thus, in practice, the users may need to be involved in the loop for making data permission decisions (elaborated in Section~\ref{sec:conclusion}).
Moreover, as we find (from Section~\ref{subsec:learn_preferences-intra-domain}) that users express varying privacy concerns across domains, custom confidence thresholds could be configured at the granularity of individual users and/or domains.
We also find (from Section~\ref{subsec:learn_preferences-inter-domain}) that certain data types show a high degree of variance across users, so they may also be excluded from predictions to avoid mistakes.

\subsubsection{Impact of permission decision history}
\label{subsec:decision_history}
In a real-world setting, users may not be expected to provide permission preferences for a large number of data types.
Thus, we explore the relation between number of training samples available per user and accuracy of our permission assistant.

\begin{table}[t]
\centering
\setlength{\tabcolsep}{2.9pt}
\begin{tabular}{lccccc}
\toprule
Hist. ratio & 0\%        & 25\%       & 50\%       & 75\%       & 100\%               \\ \midrule
\#Queries      & 0        & 1-4      & 2-8      & 3-12     & 4-16              \\ \midrule
Accuracy (\%)    & 66.9±1.6 & 77.7±1.3 & 82.1±0.7 & 83.7±0.8 & \textbf{85.1±0.4} \\
Precision (\%)   & 83.0±1.4 & 87.6±1.2 & 90.0±0.9 & 91.1±1.0 & \textbf{92.8±0.7} \\
Recall (\%)      & 65.7±0.7 & 79.1±1.5 & 83.6±0.6 & 84.7±0.7 & \textbf{85.2±1.2} \\
F1 score (\%)    & 73.4±0.9 & 83.1±1.0 & 86.6±0.4 & 87.8±0.7 & \textbf{88.8±0.5} \\
FPR (\%)         & 30.6±3.8 & 25.4±3.2 & 21.2±2.8 & 18.7±2.5 & \textbf{15.2±2.1} \\
FNR (\%)         & 34.3±0.7 & 20.9±1.5 & 16.4±0.6 & 15.3±0.7 & \textbf{14.8±1.2} \\ \bottomrule
\end{tabular}
\caption{Impact of permission history on classification.} %
\label{tab:hist_metrics}
\end{table}

To assess the impact of permission decision history on prediction accuracy, we incorporate different ratios of the permission decision history into each user's model context and evaluate the corresponding performance.
Table~\ref{tab:hist_metrics} shows the metrics for our hybrid model under various ratios of permission decision history.
We note that even without any permission history, the model still infers users' preferences by leveraging users' demographic information, AI usage experience, and privacy concerns.
We also observe that the transition from no permission history to even a small amount (i.e., permission history on 1--4 queries) yields significant improvements (i.e., more than 10.8\% increase in accuracy). 

Overall, our results indicate that an increase in permission decision history (i.e., an increase in training data) results in improved prediction accuracy.
This continuous learning capability can facilitate real-world deployment, i.e., as users naturally generate more permission history over time, the model can iteratively learn from them, leading to increasingly accurate permission predictions.

\subsection{Result Analysis}
\subsubsection{Per-user performance analysis}
\label{subsubsection:Per-UserPerformanceAnalysis}
As each user has their personalized model for predicting permission decisions, we analyze how these models perform on permission decision predictions for each user.
Specifically, we compute per-user metrics using the prediction results from each user's dedicated hybrid IC and CF model, and attempt to identify the factors that contribute to lower or higher accuracy.

\vspace{1mm}\noindent\textbf{Overall Trends.}
Figure~\ref{fig:cdf_all_metrics} presents the distribution for performance metrics computed over individual users.
Notably, 35.4\% of users achieve accuracy greater than 90\%, and 12.7\% of users even achieve perfect accuracy, meaning every permission decision predicted by their dedicated model is correct.
For precision, 70.4\% of users reach values greater than or equal to 90\%, suggesting that these models are reliable in correctly predicting data sharing (i.e., predicting positive cases).
In contrast, 34.8\% of users achieve recall values of 90\% or higher, indicating that models are generally conservative in making automated data sharing predictions. %

The error metrics further elaborate on model performance.
The distribution for FPR shows that 48.2\% of users' false permission grants (i.e., FPR) are at or below 10\% (with 0 false permission grants for 37.2\% of users).
Meanwhile, 34.8\% of users have an FNR of 10\% or less, indicating that more than a third of the users' models are able to constrain the number of false permission denials.
Overall, the personalized models show promising performance in predicting individual users' permission decisions.
In the remainder of this section, we analyze the errors to gather insights for further improving the models.

\vspace{1mm}\noindent\textbf{Detailed Analysis.}
We perform an in-depth analysis of the bottom 20\% (36) users with the lowest accuracy by comparing this group to the top 20\% of users with the highest accuracy.
There are significant differences in the average performance metrics between these two groups, for instance, average accuracy (97.6\% vs. 69.7\%), recall (98.2\% vs. 62.3\%), and precision (98.0\% vs. 69.2\%).
Next, we investigate the factors that may contribute to these differences.
We first examine whether users' self-reported attributes show significant discrepancies.
While some trends are observable, for example, high-accuracy users report slightly lower AI trust levels (2.39 vs. 2.81 on average), higher AI usage frequency (3.19 vs. 3.08), and lower privacy concerns (4.06 vs. 4.14), these differences are not statistically significant (Mann-Whitney U test $p$-values range from 0.07 to 0.95).
We then examine whether there are significant differences in the number of permission decision history entries and the number of collaborative filtering recommendations (as we observe them to be key factors in improving prediction accuracy in Section~\ref{paragraph:classification-accuracy} and~\ref{subsec:decision_history}).
We observe notable differences in the average number of permission history queries between the two groups (11.69 for high-accuracy users vs. 9.38 for low-accuracy users, U-test $p = 0.01$), as well as in the number of CF recommendations (3.96 vs. 2.90, U-test $p = 0.07$).

Next, we analyze the standard deviation in permission allowance rates across domains for both high-accuracy and low-accuracy users.
We note that high-accuracy users tend to have lower variability in their permission preferences across domains, with a larger proportion of users exhibiting smaller standard deviation values, indicating more consistent preferences.
In contrast, low-accuracy users show greater variability in their permission preferences across domains.
For instance, 30.6\% of high-accuracy users fall at the lower end of the standard deviation range ($\leq$ 0.04), compared to just 11.1\% of low-accuracy users.
Similarly, at the upper end ($\geq$ 0.27), 22.2\% of low-accuracy users exceed this standard deviation threshold (0.27), while only 13.9\% of high-accuracy users do.
We observe that high standard deviation in users' permission decisions may make it more challenging to accurately predict their permission preferences.
For instance, predicting the preferences of a user with a standard deviation of 0.43 yields subpar performance, with an accuracy of 49.0\%. %

A closer look at this user's prediction results reveals the impact of such variance: although the user is generally willing to share data in domains like \textit{Smart Home}, \textit{Travel}, and \textit{Health \& Fitness}, they tend to deny data sharing in \textit{Social}, \textit{Finance}, and \textit{Shopping}.
We surmise that when the model sees more permission history from the first group of domains, it tends to infer a higher allowance rate even for the latter group, and vice versa, during five-fold cross-validation.
This suggests that for users with high variance in their permission preferences, patterns learned from some domains may not transfer well to others, resulting in reduced prediction performance.
In contrast, when users have more consistent permission preferences across domains (i.e., lower variance), the model finds it easier to transfer learned patterns.
For example, a high-accuracy user with a standard deviation of just 0.01 achieves an accuracy of 98.0\%, recall of 100.0\%, and precision of 97.8\%.
We note that it is a key challenge to make predictions of users' permission preferences when there is a high degree of variance in their prior permission decisions.

Finally, we investigate the impact of CF recommendation accuracy on the two user groups.
We find that high-accuracy users have CF recommendations with an accuracy of 99.5\%, compared to 89.5\% CF recommendation accuracy for low-accuracy users.
Additionally, when CF recommendations are correct, the model consistently follows them, showing 100\% alignment for low-accuracy users and 99.8\% for high-accuracy users.
However, when CF recommendations are incorrect, the model still tends to follow them.
This highlights the importance of filtering out unreliable CF results to avoid misleading the model's final decisions.
For CF predictions to improve, it is crucial that users can be grouped with others who share similar preferences.
Thus, as more data is collected—particularly from a larger and more diverse user base—the quality of CF predictions naturally improves.

\subsubsection{Contextual performance analysis}
Next, we analyze how well the model predicts user preferences under different contexts, specifically across domains, tools, and data types.
Our goal is to examine whether there are significant differences in accuracy, FPR, and FNR among these contextual factors, thereby revealing patterns in when and where the model tends to be over- or under-permissive.

\vspace{1mm}\noindent\textbf{Performance by Domain.}
When analyzing performance across different domains, we find that the model performs slightly better at predicting permission requests in certain domains.
Notably, \textit{Work \& Productivity} (87.4\% accuracy), \textit{Health \& Fitness} (87.1\%), and \textit{Entertainment} (86.8\%) exhibit relatively high accuracies.
In contrast, domains such as \textit{Shopping} (84.3\% accuracy), \textit{Finance} (81.5\%), and \textit{Smart Home} (80.1\%) show lower prediction accuracies.

We find that these results correlate with the variance of permission decisions in domains (Figure~\ref{fig:variance_domain} plots standard deviation across domains). 
Specifically, the high-accuracy domains have lower variance in participants' permission preferences and low-accuracy domains have the highest variance in participants' permission preferences.
A closer examination of the \textit{Smart Home} results reveals a high FNR of 23.8\%, indicating a conservative bias in which the model tends to reject permissions that users might have accepted in this domain.
Notably, \textit{Finance} has the highest FNR of all domains, at 28.3\%, which reflects the model's cautious behavior in this particularly sensitive domain.
By contrast, the \textit{Social} domain exhibits the highest FPR, at 23.3\%, suggesting that the model is more likely to approve permission requests in less sensitive domains.

Overall, we note that while variance within domains degrades accuracy, the degradation is not substantial, especially as compared to the degradation in accuracy with variance within the user's permission decisions (Section~\ref{subsubsection:Per-UserPerformanceAnalysis}).

\vspace{1mm}\noindent\textbf{Performance by Tool.}
At the tool level, disparities become even more apparent.
Tools with the highest accuracies include \textit{Calendar} (92.0\%), \textit{The Movie Database} (89.0\%), \textit{Tinder} (88.8\%), \textit{Hospital Booking} (88.6\%), and \textit{Weather} (88.4\%).
These are also the tools for which participants tend to grant more permissions.
For example, as shown in Table~\ref{tab:tool_analysis}, \textit{Calendar} appears in multiple domains such as \textit{Travel}, \textit{Work \& Productivity}, and \textit{Health \& Fitness}, and it ranks among the top tools with the most \textit{Always allow} permissions.
A similar pattern holds for the other aforementioned tools, suggesting that the model predicts permissions more accurately when there is more data available, and users have consistent preferences for specific tools.

In contrast, some tools have noticeably lower accuracies, including \textit{SMS} (76.4\%), \textit{Cloud Drive} (76.4\%), \textit{Slack} (79.3\%), \textit{Lighting Control} (80.0\%), and \textit{Banking} (80.3\%).
Among these tools, \textit{Cloud Drive} (FNR of 31.4\%), \textit{Banking} (27.3\%), and \textit{Lighting Control} (20.9\%) suffer from high FNRs, reflecting the model's tendency toward caution.

\vspace{1mm}\noindent\textbf{Performance by Data Type.}
We observe substantial differences while analyzing the model's accuracy across data types. 
For instance, some data types yield extremely high accuracies, even reaching 100\%, such as \textit{Fitness Goal} and \textit{Hobbies and Interests}. Other data types with similarly strong results are \textit{Relationship Preferences} (98.1\%), \textit{Personal Biography} (96.2\%), \textit{Travel Destination} (95.5\%), and \textit{Accommodation Details} (95.4\%).
It is worth noting that these data types also rank among those with the lowest preference variance among participants, as shown in Table~\ref{tab:10_datatype_std}. This supports our hypothesis that data types with more consistent user preferences are easier for the model to predict accurately.

On the other hand, data types like \textit{Payment Method Detail} (59.1\%), \textit{Employment Details} (63.6\%), \textit{Driver License Number} (71.4\%), and \textit{Travel Itinerary} (75.9\%) show some of the lowest accuracies.
These data types also have the highest variance in user preferences, according to Table~\ref{tab:10_datatype_std}. In other words, data types with more diverse permission preferences tend to be more difficult for the model to predict.
Other low-accuracy data types often involve highly sensitive information as well, such as \textit{Personal Identification Numbers} (61.9\%) and \textit{Travel History} (63.2\%).
These types of data also tend to have higher FNRs.
For instance, \textit{Payment Method Details} has an FNR of 87.5\%, \textit{Personal Identification Numbers} has 72.2\%, and \textit{Driver License Number} has 66.7\%. These are among the most under-permissioned data types. Only a few show relatively high FPRs, such as \textit{Employment Details} and \textit{Travel History}, both at 28.6\%.
These results suggest that the model often misclassifies these sensitive data types as unacceptable to share, even when users may permit it. This reflects challenges in accurately capturing context-sensitive boundaries for personally identifiable or other sensitive information.

%% file: 6_discussion.tex
\section{Discussion}

\vspace{1mm}
\noindent
\textbf{Reinforcing and Expanding Prior Knowledge.}
Our findings on user permissions for AI agents confirm established access control principles from traditional systems while also revealing agent-specific dynamics. Consistent with prior work on mobile applications and voice assistants\cite{abdi2021privacy,lin2014modeling,malkin2022runtime,peddinti2019reducing,seneviratne2015your}, our results show that over-permissioning remains a key challenge, and that user decisions are highly dependent on context, such as data sensitivity and the information's recipient. However, the autonomous nature of AI agents fundamentally alters the control dynamic. We find that permission granting becomes a continuous process of trust evaluation: when an agent makes a mistake, users are significantly less willing to grant permissions, indicating that agent performance is a new, critical factor influencing privilege. Most critically for our work, we find that despite these complexities, individual preferences exhibit a high degree of consistency within specific contexts and across users. This predictable pattern is the key insight that enables our permission-prediction system, demonstrating that foundational privacy norms persist while the agent's autonomy introduces new factors for permission management.

\vspace{1mm}
\noindent
\textbf{Model Robustness and Limitations.}
As agents interface with data and resources from potentially untrustworthy entities, it is crucial that the permission inference is robust against such attacks.
When users communicate with AI agents using natural language instructions, they often reveal implicit or explicit preferences about what the agent is allowed to do.
Such information can complement the execution context and past user behavior in making accurate predictions.
In fact, prior work has used such information to predict the expected control and data flow of AI agents and detect anomalies~\cite{debenedetti2025defeating, wu2025isolategpt}.
Similarly, prior work has proposed AI agent architectures that, by design, limit the flow of information between system modules~\cite{bagdasarian2024airgapagent, wu2025isolategpt} and reliably control the generation of text in LLMs~\cite{Moskal2024}.
We believe that such approaches can be extended to support robust permission management in AI agents.

That said, natural language interactions can be ambiguous, and this remains an open problem for LLMs~\cite{liu2023LLMAmbguity}.
Our in-context learning model can be affected by this ambiguity and may make mistakes when processing unclear instructions.
However, because we also rely on collaborative filtering over deterministic data types, the system inherently offers some resilience to such ambiguity for previously seen data types.
Once a request is mapped to a canonical label, collaborative filtering predictions become insensitive to wording.
Moreover, taking only high-confidence predictions and delegating uncertain data access permissions to users can help mitigate the model's robustness issues.
This approach pairs predictions with clear controls to make and revoke decisions and to provide feedback, which reduces interruptions for routine cases while allowing human oversight in riskier or uncertain situations.

\vspace{1mm}\noindent
\textbf{Towards Usable Permission Management.}
Automated permission management is a multifaceted problem. 
Our work in this paper focuses on one facet, and other important facets, including improving the usability of permission management in AI agents, remain challenges and open avenues for future work.
For example, during early use, the assistant may still need to learn a user's preferences; a user's preferences may change as their circumstances change (e.g., a previously trusted entity becomes untrusted); and some situations may be fundamentally difficult to predict.
Thus, a full-fledged permission system will need to include not only a permission prediction module, but also UI/UX for engaging the user directly in cases where the predictive model has low confidence or makes a mistake---for example, UI/UX for users to make explicit permission decisions, to revoke previously-granted permissions, and to give feedback to the permission assistant.
We believe there is rich future work to be done on how to design such a hybrid system well. 
The important contribution of our work here is to start this line of inquiry and to demonstrate that it is feasible: an effective AI permission assistant can substantially reduce the number of decisions that users must be asked to make or evaluate directly, paving the way for a usable and secure permission management system.

%% file: 7_conclusion.tex
\section{Conclusion}
\label{sec:conclusion}

In this paper, we explored the capabilities and limits of automating permission management in AI agents. Through a user study, we found that long-standing challenges such as over-permissioning persist in agentic systems, alongside new issues unique to them. Notably, when agents make mistakes, users become less willing to grant permissions, indicating that agent performance is a critical factor influencing privilege decisions. We also found that permission choices are strongly shaped by communication context, yet remain consistent within that context across user groups.
Leveraging these insights, we developed a permission prediction model that combines collaborative filtering with LLM in-context learning, achieving 94.4\% accuracy for high-confidence predictions. Even without prior permission history, the model reached 66.9\% accuracy, with just 1–4 samples improving it by 10.8\%.
Our findings demonstrate the potential of learning user preferences to automate many permission decisions. However, challenges remain in enforcing predictions, improving robustness, and designing usable interfaces. Overall, this work advances automated permission management and outlines key directions toward its practical deployment.

%% file: acknowledgment.tex
\section*{Acknowledgment}
We thank the reviewers for their valuable feedback.
This work was partially supported by NSF (CNS-2154930, CNS-2238635),
ARO (W911NF-24-1-0155), ONR (N000142412663), and gifts from Microsoft. 
T.~Kohno was also supported by the McDevitt Chair in Computer Science, Ethics, and Society at Georgetown University; the majority of this research was conducted while he was at the University of Washington.
We would also like to thank Pardis Emami-Naeini for providing feedback on the user study and statistical analysis.

%% file: X_appendix.tex
\section{Analysis of Demographics and Self-Reported Metrics}
\label{sec:ana_demo}

\begin{figure}[t]
\centering
\begin{subfigure}[b]{0.16\textwidth}
\centering
    \includegraphics[height=1.6cm]{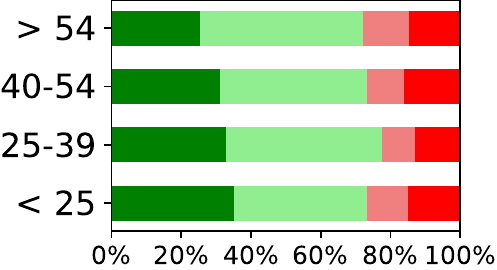}
    \caption{Age group\label{fig:age}}
\end{subfigure}
\begin{subfigure}[b]{0.16\textwidth}
\centering
    \includegraphics[height=1.6cm]{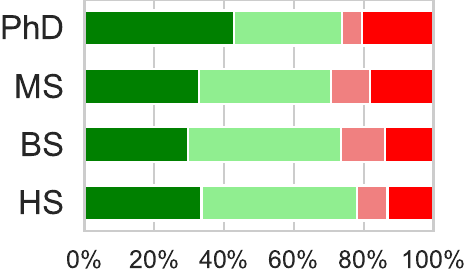}
    \caption{Education level\label{fig:education}}
\end{subfigure}
\begin{subfigure}[b]{0.145\textwidth}
\centering
    \includegraphics[height=1.6cm]{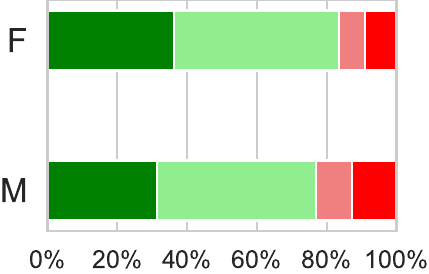}
    \caption{Sex}
\end{subfigure}
\vspace{0.3cm}
\begin{subfigure}[b]{0.225\textwidth}
\includegraphics[width=\textwidth]{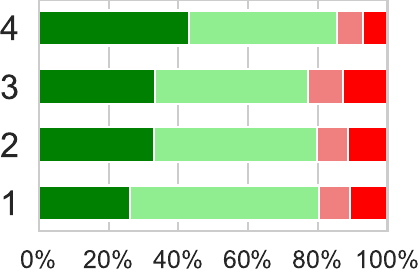}
\caption{AI tool familiarity\label{fig:ai-familiarity}}
\end{subfigure}
\begin{subfigure}[b]{0.225\textwidth}
\includegraphics[width=\textwidth]{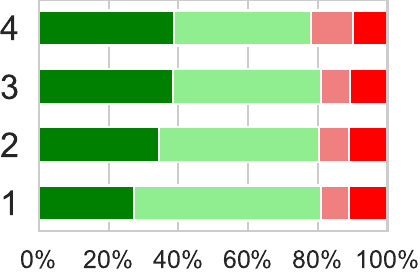}
\caption{AI tool usage frequency\label{fig:ai_frequency}}
\end{subfigure}
\vspace{0.3cm}
\begin{subfigure}[b]{0.225\textwidth}
\includegraphics[width=\textwidth]{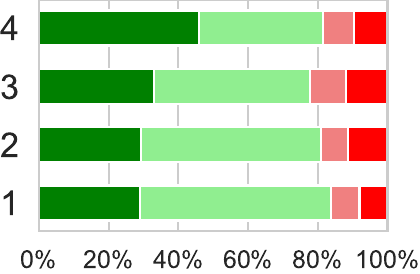}
\caption{AI tool trust\label{fig:ai_trust}}
\end{subfigure}
\begin{subfigure}[b]{0.225\textwidth}
\includegraphics[width=\textwidth]{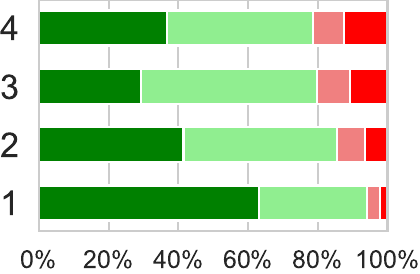}
\caption{Privacy consciousness\label{fig:privacy_consciousness}}
\end{subfigure}
\caption{Group users according to their demographic information, which includes age group, education level, and sex, as well as their self-reported metrics such as AI tool familiarity, AI tool usage frequency, AI tool trust, and privacy consciousness level. For age groups, participants are divided into \textit{below 25}, \textit{25-39}, \textit{40-55}, and \textit{over 55}. Regarding education level, participants are categorized into four groups: high school or less (\textit{HS}), bachelor’s (\textit{BS}), master’s (\textit{MS}), and PhD or other doctorate (\textit{PhD}). For sex, \textit{F} denotes female and \textit{M} denotes male. Self-reported metrics are measured on a four-level scale (1-4) where a higher number indicates a greater level; for example, a privacy consciousness level of 4 signifies the highest concern for privacy.} %
\label{fig:demo_info}
\end{figure}

We analyze each participant’s basic demographic information and self-reported metrics related to AI usage and privacy consciousness to investigate whether these factors significantly influence their permission preferences (see Figure~\ref{fig:demo_info}).
Figure~\ref{fig:age} shows that participants under the age of 25 selected \textit{Always Allow} permissions more frequently than those over 55 (35.0\% vs. 25.5\%), suggesting a generational difference in privacy attitudes.
Additionally, participants with higher levels of education granted fewer data-sharing permissions overall (Figure~\ref{fig:education}).

The self-reported metrics include AI tool familiarity, usage frequency, trust in AI tools, and privacy consciousness. We observe a significant shift in permission preference distributions across different levels of these metrics.
As shown in Figure~\ref{fig:ai-familiarity}, participants more familiar with AI tools were more likely to select \textit{Always Allow} and less likely to choose \textit{Never Share}.
Although AI usage frequency and trust (Figures~\ref{fig:ai_frequency} and~\ref{fig:ai_trust}) do not show strong effects on overall permission preferences, those who used AI tools more often and reported higher trust levels tended to favor \textit{Always Allow}. This suggests that greater engagement with AI may correlate with a preference for smoother user experiences, enabled by more permissive data-sharing settings.

For the privacy consciousness (Figure~\ref{fig:privacy_consciousness}), participants with higher levels of privacy consciousness were more likely to choose \textit{Never Share}, reflecting a stronger desire to restrict data sharing.
Interestingly, participants with medium levels of privacy consciousness selected \textit{Always Allow} less often than those with high levels, but overall, they exhibited a more permissive data-sharing pattern compared to highly privacy-conscious individuals.
While these self-reported measures may lack standardization or objectivity~\cite{chan2010so}, and some participants may have inaccurate perceptions of their privacy attitudes, the overall trends provide useful insights into potential user data-sharing behaviors.
In summary, our analysis indicates that demographic factors, AI-related attitudes, and privacy consciousness are associated with participants’ data-sharing preferences. These findings help us identify patterns that can be used to anticipate general user attitudes toward data-sharing permissions.

\begin{figure}[t]
    \centering
    \includegraphics[width=0.465\textwidth]{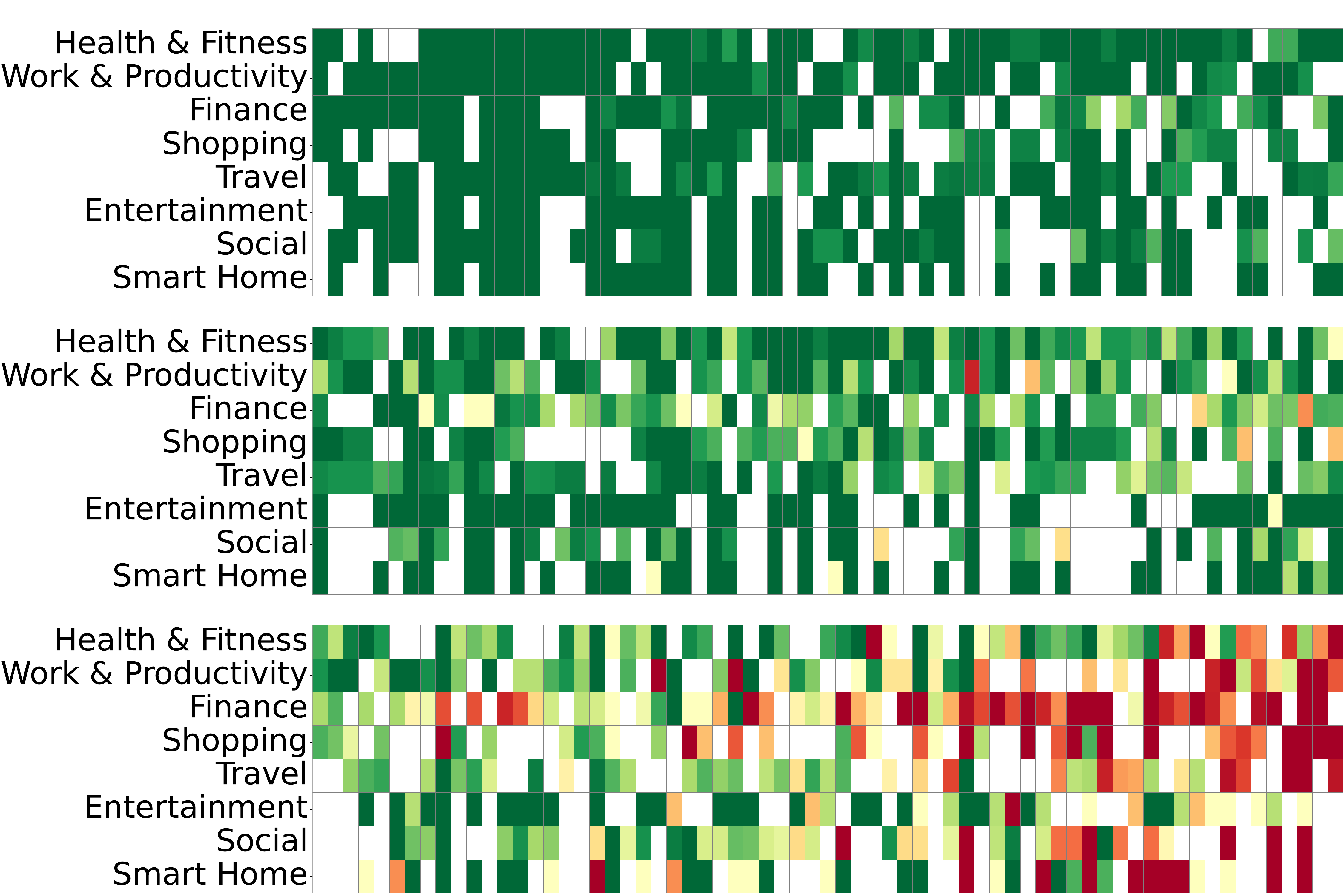}
    \caption{Heatmap of permission allowance rates across various domains for individual users. The figure is segmented into three distinct user groups. In each segmentation (user group), each column corresponds to a user and each row to a distinct domain (\textit{Health \& Fitness}, \textit{Work \& Productivity}, \textit{Finance}, \textit{Shopping}, \textit{Travel}, \textit{Entertainment}, \textit{Social}, \textit{Smart Home}). Darker green indicates a higher permission allowance rate, while darker red indicates a lower rate.}
    \label{fig:heatmap_stacked}
\end{figure}

\section{Analysis of User Permission Preferences Across Domains}
\label{sec:heatmap}
Figure~\ref{fig:heatmap_stacked} provides insights into data-sharing permission preferences across domains among participants.
It reveals a high degree of consistency in overall permission behaviors, as many participants exhibit uniform preferences across different domains, evident from extensive areas of homogeneous coloring.
At the same time, variations in color intensity highlight specific domains, such as \textit{Finance} and \textit{Shopping}, where permission preferences significantly differ from more permissive areas like \textit{Health \& Fitness} and \textit{Social} for some participants.
These patterns emphasize that while participants generally maintain consistent permission preferences across domains, the distinct contexts of each domain can lead to observable variations in the data-sharing preferences of some participants.

\begin{figure}[t]
    \centering
    \includegraphics[width=0.9\linewidth]{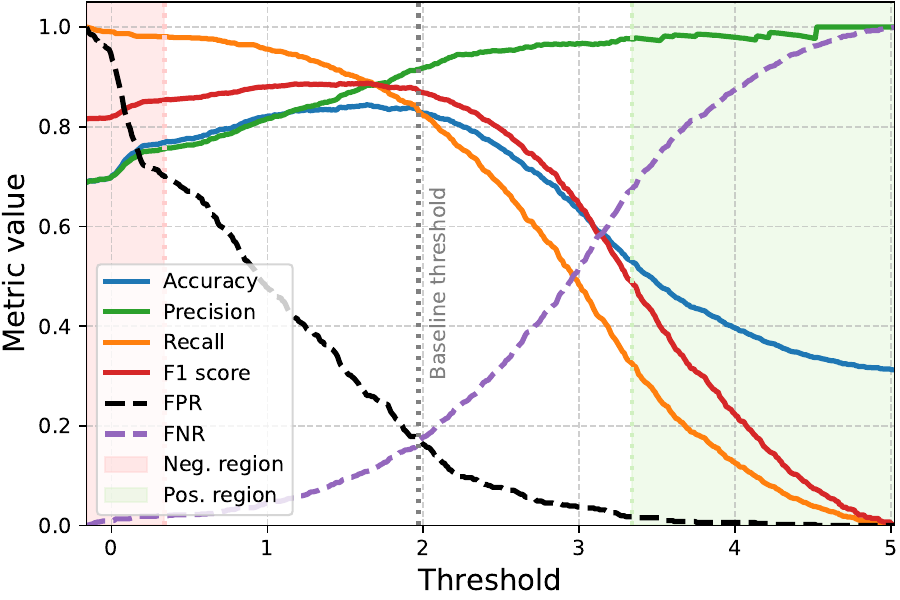}
    \caption{Collaborative filtering model metrics under different prediction score thresholds.}
    \label{fig:cf_metrics_vs_prediction_threshold}
\end{figure}

\section{CF Model Prediction Threshold}
\label{sec:cf_raw_metrics}
Since the CF model's performance is highly sensitive to the prediction score threshold, we establish a baseline threshold for each fold in five-fold cross-validation by selecting the point where the FPR equals the FNR.
Figure~\ref{fig:cf_metrics_vs_prediction_threshold} illustrates the CF model's metrics under varying thresholds for one of our test runs and shows the selected baseline threshold for reporting the model comparison results.
For integrating the IC with CF, we use only CF predictions with high confidence.
In the figure, the \textit{negative region} and \textit{positive region} define ranges of prediction scores that serve as additional contextual examples for the hybrid model.

%% file: X_meta_review.tex
\section{Meta-Review}

The following meta-review was prepared by the program committee for the 2026
IEEE Symposium on Security and Privacy (S\&P) as part of the review process as
detailed in the call for papers.

\subsection{Summary}
This paper presents an automated permission management system for AI agents. The authors conducted a vignette-based survey of 205 users across a range of domains and scenarios to understand how users make permission decisions regarding data access and sharing in agentic systems. Building on these findings, the paper presents a preliminary design of a permission prediction model.

\subsection{Scientific Contributions}
\begin{itemize}
\item Independent Confirmation of Important Results with Limited Prior Research
\item Establishes a New Research Direction
\end{itemize}

\subsection{Reasons for Acceptance}
\begin{enumerate}
\item The paper builds on a long line of work that seeks to understand factors influencing users' decisions in terms of sharing data and granting permissions to automated systems. Prior work in this space has focused on non-LLM-based personal assistants, and this paper obtains results for users' preferences in LLM-based agentic systems.
\item This paper is one of the first to investigate the problem of automating permission management for AI agents. It argues that, since automation is a key value proposition of AI agents, there is a need for permission management systems that can automatically make decisions on user’s behalf. The paper establishes this new subfield of automated permission systems for AI agents, and presents one point in the design space of this problem.
\end{enumerate}

%% file: main.bbl
\begin{thebibliography}{10}

\bibitem{openai-gpt-action-authentication}
Gpt action authentication.
\newblock \url{https://platform.openai.com/docs/actions/authentication}, 2025.
\newblock Accessed: 2025-06-02.

\bibitem{abdi2021privacy}
Noura Abdi, Xiao Zhan, Kopo~M Ramokapane, and Jose Such.
\newblock Privacy norms for smart home personal assistants.
\newblock In {\em Proceedings of the 2021 CHI conference on human factors in computing systems}, pages 1--14, 2021.

\bibitem{almuhimedi2015your}
Hazim Almuhimedi, Florian Schaub, Norman Sadeh, Idris Adjerid, Alessandro Acquisti, Joshua Gluck, Lorrie~Faith Cranor, and Yuvraj Agarwal.
\newblock Your location has been shared 5,398 times! a field study on mobile app privacy nudging.
\newblock In {\em Proceedings of the 33rd annual ACM conference on human factors in computing systems}, pages 787--796, 2015.

\bibitem{amoros2023predicting}
Marc~Serramia Amoros, William Seymour, Natalia Criado, and Michael Luck.
\newblock Predicting privacy preferences for smart devices as norms.
\newblock In {\em The 22nd International Conference on Autonomous Agents and Multiagent Systems}. International Foundation for Autonomous Agents and Multiagent Systems (IFAAMAS), 2023.

\bibitem{AndroidPermissions}
{Android Developers}.
\newblock Permissions overview.
\newblock \url{https://developer.android.com/guide/topics/permissions/overview}, 2023.
\newblock Accessed: 2025-03-26.

\bibitem{AndroidRuntimePerms}
{Android Open Source Project}.
\newblock Runtime permissions.
\newblock \url{https://source.android.com/docs/core/permissions/runtime_perms}, 2023.
\newblock Accessed: 2025-03-26.

\bibitem{anthropic_computer_use_2024}
{Anthropic}.
\newblock Developing a computer use model, 2024.

\bibitem{bagdasarian2024airgapagent}
Eugene Bagdasarian, Ren Yi, Sahra Ghalebikesabi, Peter Kairouz, Marco Gruteser, Sewoong Oh, Borja Balle, and Daniel Ramage.
\newblock Airgapagent: Protecting privacy-conscious conversational agents.
\newblock In {\em Proceedings of the 2024 on ACM SIGSAC Conference on Computer and Communications Security}, pages 3868--3882, 2024.

\bibitem{barbosa2019if}
Nat{\~a}~M Barbosa, Joon~S Park, Yaxing Yao, and Yang Wang.
\newblock “what if?” predicting individual users’ smart home privacy preferences and their changes.
\newblock {\em Proceedings on Privacy Enhancing Technologies}, 2019.

\bibitem{barth2006privacy}
Adam Barth, Anupam Datta, John~C Mitchell, and Helen Nissenbaum.
\newblock Privacy and contextual integrity: Framework and applications.
\newblock In {\em 2006 IEEE symposium on security and privacy (S\&P'06)}, pages 15--pp. IEEE, 2006.

\bibitem{bilogrevic2021shhh}
Igor Bilogrevic, Balazs Engedy, Judson~L Porter~III, Nina Taft, Kamila Hasanbega, Andrew Paseltiner, Hwi~Kyoung Lee, Edward Jung, Meggyn Watkins, PJ~McLachlan, et~al.
\newblock " shhh... be $\{$quiet!$\}$" reducing the unwanted interruptions of notification permission prompts on chrome.
\newblock In {\em 30th USENIX Security Symposium (USENIX Security 21)}, pages 769--784, 2021.

\bibitem{brandao2022prediction}
Andr{\'e} Brand{\~a}o, Ricardo Mendes, and Jo{\~a}o~P Vilela.
\newblock Prediction of mobile app privacy preferences with user profiles via federated learning.
\newblock In {\em Proceedings of the Twelfth ACM Conference on Data and Application Security and Privacy}, pages 89--100, 2022.

\bibitem{brown2020language}
Tom Brown, Benjamin Mann, Nick Ryder, Melanie Subbiah, Jared~D Kaplan, Prafulla Dhariwal, Arvind Neelakantan, Pranav Shyam, Girish Sastry, Amanda Askell, et~al.
\newblock Language models are few-shot learners.
\newblock {\em Advances in neural information processing systems}, 33:1877--1901, 2020.

\bibitem{chan2010so}
David Chan.
\newblock So why ask me? are self-report data really that bad?
\newblock In {\em Statistical and methodological myths and urban legends}, pages 329--356. Routledge, 2010.

\bibitem{das2018personalized}
Anupam Das, Martin Degeling, Daniel Smullen, and Norman Sadeh.
\newblock Personalized privacy assistants for the internet of things: Providing users with notice and choice.
\newblock {\em IEEE Pervasive Computing}, 17(3):35--46, 2018.

\bibitem{debenedetti2025defeating}
Edoardo Debenedetti, Ilia Shumailov, Tianqi Fan, Jamie Hayes, Nicholas Carlini, Daniel Fabian, Christoph Kern, Chongyang Shi, Andreas Terzis, and Florian Tram{\`e}r.
\newblock Defeating prompt injections by design.
\newblock {\em arXiv preprint arXiv:2503.18813}, 2025.

\bibitem{debenedetti2024agentdojo}
Edoardo Debenedetti, Jie Zhang, Mislav Balunovi{\'c}, Luca Beurer-Kellner, Marc Fischer, and Florian Tram{\`e}r.
\newblock Agentdojo: A dynamic environment to evaluate attacks and defenses for llm agents.
\newblock {\em arXiv preprint arXiv:2406.13352}, 2024.

\bibitem{distler2023empirical}
Verena Distler, Matthias Fassl, Hana Habib, Katharina Krombholz, Gabriele Lenzini, Carine Lallemand, Vincent Koenig, and Lorrie~Faith Cranor.
\newblock Empirical research methods in usable privacy and security.
\newblock In {\em Human Factors in Privacy Research}, pages 29--53. Springer International Publishing Cham, 2023.

\bibitem{dong2024survey}
Qingxiu Dong, Lei Li, Damai Dai, Ce~Zheng, Jingyuan Ma, Rui Li, Heming Xia, Jingjing Xu, Zhiyong Wu, Baobao Chang, et~al.
\newblock A survey on in-context learning.
\newblock In {\em Proceedings of the 2024 Conference on Empirical Methods in Natural Language Processing}, pages 1107--1128, 2024.

\bibitem{EnglehardtOpenWPMMeasurement}
Steven Englehardt and Arvind Narayanan.
\newblock Online tracking: A 1-million-site measurement and analysis.
\newblock In {\em Proceedings of the 2016 ACM SIGSAC conference on computer and communications security}, pages 1388--1401, 2016.

\bibitem{felt2011android}
Adrienne~Porter Felt, Erika Chin, Steve Hanna, Dawn Song, and David Wagner.
\newblock Android permissions demystified.
\newblock In {\em Proceedings of the 18th ACM conference on Computer and communications security}, pages 627--638, 2011.

\bibitem{felt2012android}
Adrienne~Porter Felt, Elizabeth Ha, Serge Egelman, Ariel Haney, Erika Chin, and David Wagner.
\newblock Android permissions: User attention, comprehension, and behavior.
\newblock In {\em Proceedings of the eighth symposium on usable privacy and security}, pages 1--14, 2012.

\bibitem{filipczuk2022automated}
Dorota Filipczuk, Tim Baarslag, Enrico~H Gerding, and MC~Schraefel.
\newblock Automated privacy negotiations with preference uncertainty.
\newblock {\em Autonomous Agents and Multi-Agent Systems}, 36(2):49, 2022.

\bibitem{finch1987vignette}
Janet Finch.
\newblock The vignette technique in survey research.
\newblock {\em Sociology}, 21(1):105--114, 1987.

\bibitem{google_project_mariner_2024}
{Google DeepMind}.
\newblock Introducing gemini 2.0: our new ai model for the agentic era, 2024.

\bibitem{harbach2024don}
Marian Harbach, Igor Bilogrevic, Enrico Bacis, Serena Chen, Ravjit Uppal, Andy Paicu, Elias Klim, Meggyn Watkins, and Balazs Engedy.
\newblock Don’t interrupt me--a large-scale study of on-device permission prompt quieting in chrome.
\newblock In {\em NDSS. The Internet Society, San Diego, CA}, 2024.

\bibitem{harborth2021investigating}
David Harborth and Sebastian Pape.
\newblock Investigating privacy concerns related to mobile augmented reality apps--a vignette based online experiment.
\newblock {\em Computers in Human Behavior}, 122:106833, 2021.

\bibitem{langchain}
{Harrison Chase and the LangChain Team}.
\newblock Langchain.
\newblock \url{https://www.langchain.com/}, 2025.

\bibitem{he2020lightgcn}
Xiangnan He, Kuan Deng, Xiang Wang, Yan Li, Yongdong Zhang, and Meng Wang.
\newblock Lightgcn: Simplifying and powering graph convolution network for recommendation.
\newblock In {\em Proceedings of the 43rd International ACM SIGIR conference on research and development in Information Retrieval}, pages 639--648, 2020.

\bibitem{he2017neural}
Xiangnan He, Lizi Liao, Hanwang Zhang, Liqiang Nie, Xia Hu, and Tat-Seng Chua.
\newblock Neural collaborative filtering.
\newblock In {\em Proceedings of the 26th international conference on world wide web}, pages 173--182, 2017.

\bibitem{huang2025survey}
Lei Huang, Weijiang Yu, Weitao Ma, Weihong Zhong, Zhangyin Feng, Haotian Wang, Qianglong Chen, Weihua Peng, Xiaocheng Feng, Bing Qin, et~al.
\newblock A survey on hallucination in large language models: Principles, taxonomy, challenges, and open questions.
\newblock {\em ACM Transactions on Information Systems}, 43(2):1--55, 2025.

\bibitem{iqbal2022your}
Umar Iqbal, Pouneh~Nikkhah Bahrami, Rahmadi Trimananda, Hao Cui, Alexander Gamero-Garrido, Daniel Dubois, David Choffnes, Athina Markopoulou, Franziska Roesner, and Zubair Shafiq.
\newblock Tracking, profiling, and ad targeting in the alexa echo smart speaker ecosystem.
\newblock In {\em ACM Internet Measurement Conference (IMC)}, 2023.

\bibitem{iqbal2024llm}
Umar Iqbal, Tadayoshi Kohno, and Franziska Roesner.
\newblock Llm platform security: applying a systematic evaluation framework to openai's chatgpt plugins.
\newblock In {\em Proceedings of the AAAI/ACM Conference on AI, Ethics, and Society}, volume~7, pages 611--623, 2024.

\bibitem{iqbal2024llmAIES}
Umar Iqbal, Tadayoshi Kohno, and Franziska Roesner.
\newblock Llm platform security: applying a systematic evaluation framework to openai's chatgpt plugins.
\newblock In {\em Proceedings of the AAAI/ACM Conference on AI, Ethics, and Society}, volume~7, pages 611--623, 2024.

\bibitem{llamaindex}
{Jerry Liu and Simon Suo and the LlamaIndex Team}.
\newblock Llamaindex.
\newblock \url{https://www.llamaindex.ai/}, 2025.

\bibitem{jia2017contexlot}
Yunhan~Jack Jia, Qi~Alfred Chen, Shiqi Wang, Amir Rahmati, Earlence Fernandes, Zhuoqing~Morley Mao, Atul Prakash, and SJ~Unviersity.
\newblock Contexlot: Towards providing contextual integrity to appified iot platforms.
\newblock In {\em ndss}, volume~2, pages 2--2. San Diego, 2017.

\bibitem{kelley2012conundrum}
Patrick~Gage Kelley, Sunny Consolvo, Lorrie~Faith Cranor, Jaeyeon Jung, Norman Sadeh, and David Wetherall.
\newblock A conundrum of permissions: installing applications on an android smartphone.
\newblock In {\em Financial Cryptography and Data Security: FC 2012 Workshops, USEC and WECSR 2012, Kralendijk, Bonaire, March 2, 2012, Revised Selected Papers 16}, pages 68--79. Springer, 2012.

\bibitem{lassakintroducing}
Leona Lassak, Hanna P{\"u}schel, Tobias Gostomzyk, and Markus D{\"u}rmuth.
\newblock Introducing data trustees: A vignette-based study approach to get users in the loop.
\newblock 2023.

\bibitem{lassakbalancing}
Leona Lassak, Hanna P{\"u}schel, Oliver~D Reithmaier, Tobias Gostomzyk, and Markus D{\"u}rmuth.
\newblock Balancing privacy and data utilization: A comparative vignette study on user acceptance of data trustees in germany and the us.
\newblock In {\em Network and Distributed System Security (NDSS) Symposium}, 2025.

\bibitem{lau2018alexa}
Josephine Lau, Benjamin Zimmerman, and Florian Schaub.
\newblock Alexa, are you listening? privacy perceptions, concerns and privacy-seeking behaviors with smart speakers.
\newblock {\em Proceedings of the ACM on human-computer interaction}, 2(CSCW):1--31, 2018.

\bibitem{li2024dissecting}
Junlong Li, Fan Zhou, Shichao Sun, Yikai Zhang, Hai Zhao, and Pengfei Liu.
\newblock Dissecting human and llm preferences.
\newblock In {\em Proceedings of the 62nd Annual Meeting of the Association for Computational Linguistics}, pages 1790--1811, 2024.

\bibitem{li2022scenario}
Xiaotian~Vivian Li, Mary~Beth Rosson, and Jenay Robert.
\newblock A scenario-based exploration of expected usefulness, privacy concerns, and adoption likelihood of learning analytics.
\newblock In {\em Proceedings of the ninth ACM conference on learning@ scale}, pages 48--59, 2022.

\bibitem{lin2014modeling}
Jialiu Lin, Bin Liu, Norman Sadeh, and Jason~I Hong.
\newblock Modeling $\{$Users’$\}$ mobile app privacy preferences: Restoring usability in a sea of permission settings.
\newblock In {\em 10th Symposium On Usable Privacy and Security (SOUPS 2014)}, pages 199--212, 2014.

\bibitem{liu2023we}
Alisa Liu, Zhaofeng Wu, Julian Michael, Alane Suhr, Peter West, Alexander Koller, Swabha Swayamdipta, Noah~A Smith, and Yejin Choi.
\newblock We’re afraid language models aren’t modeling ambiguity.
\newblock In {\em Proceedings of the 2023 Conference on Empirical Methods in Natural Language Processing}, pages 790--807, 2023.

\bibitem{liu2023LLMAmbguity}
Alisa Liu, Zhaofeng Wu, Julian Michael, Alane Suhr, Peter West, Alexander Koller, Swabha Swayamdipta, Noah~A Smith, and Yejin Choi.
\newblock We’re afraid language models aren’t modeling ambiguity.
\newblock In {\em Proceedings of the 2023 Conference on Empirical Methods in Natural Language Processing}, pages 790--807, 2023.

\bibitem{liu2016follow}
Bin Liu, Mads~Schaarup Andersen, Florian Schaub, Hazim Almuhimedi, Shikun~Aerin Zhang, Norman Sadeh, Yuvraj Agarwal, and Alessandro Acquisti.
\newblock Follow my recommendations: A personalized privacy assistant for mobile app permissions.
\newblock In {\em Twelfth symposium on usable privacy and security (SOUPS 2016)}, pages 27--41, 2016.

\bibitem{liu2022effects}
Gary Liu and Nathan Malkin.
\newblock Effects of privacy permissions on user choices in voice assistant app stores.
\newblock {\em Proceedings on Privacy Enhancing Technologies}, 2022.

\bibitem{malkin2019privacy}
Nathan Malkin, Serge Egelman, and David Wagner.
\newblock Privacy controls for always-listening devices.
\newblock In {\em Proceedings of the New Security Paradigms Workshop}, pages 78--91, 2019.

\bibitem{malkin2022runtime}
Nathan Malkin, David Wagner, and Serge Egelman.
\newblock Runtime permissions for privacy in proactive intelligent assistants.
\newblock In {\em Eighteenth Symposium on Usable Privacy and Security (SOUPS 2022)}, pages 633--651, 2022.

\bibitem{manusai}
{Manus AI}.
\newblock Manus - an autonomous artificial intelligence agent.
\newblock \url{https://manus.im/}.

\bibitem{mendes2022enhancing}
Ricardo Mendes, Mariana Cunha, Jo{\~a}o~P Vilela, and Alastair~R Beresford.
\newblock Enhancing user privacy in mobile devices through prediction of privacy preferences.
\newblock In {\em European Symposium on Research in Computer Security}, pages 153--172. Springer, 2022.

\bibitem{Moskal2024}
Michal Moskal, Madan Musuvathi, and Emre {K\i c\i man}.
\newblock {AI Controller Interface}.
\newblock \url{https://github.com/microsoft/aici/}, 2024.

\bibitem{motiee2010windows}
Sara Motiee, Kirstie Hawkey, and Konstantin Beznosov.
\newblock Do windows users follow the principle of least privilege? investigating user account control practices.
\newblock In {\em Proceedings of the Sixth Symposium on Usable Privacy and Security}, pages 1--13, 2010.

\bibitem{movva2024annotation}
Rajiv Movva, Pang~Wei Koh, and Emma Pierson.
\newblock Annotation alignment: Comparing llm and human annotations of conversational safety.
\newblock In {\em Proceedings of the 2024 Conference on Empirical Methods in Natural Language Processing}, pages 9048--9062, 2024.

\bibitem{naeini2017privacy}
Pardis~Emami Naeini, Sruti Bhagavatula, Hana Habib, Martin Degeling, Lujo Bauer, Lorrie~Faith Cranor, and Norman Sadeh.
\newblock Privacy expectations and preferences in an $\{$IoT$\}$ world.
\newblock In {\em Thirteenth symposium on usable privacy and security (SOUPS 2017)}, pages 399--412, 2017.

\bibitem{nissenbaum2004privacy}
Helen Nissenbaum.
\newblock Privacy as contextual integrity.
\newblock {\em Wash. L. Rev.}, 79:119, 2004.

\bibitem{nori2023can}
Harsha Nori, Yin~Tat Lee, Sheng Zhang, Dean Carignan, Richard Edgar, Nicolo Fusi, Nicholas King, Jonathan Larson, Yuanzhi Li, Weishung Liu, et~al.
\newblock Can generalist foundation models outcompete special-purpose tuning? case study in medicine.
\newblock {\em arXiv preprint arXiv:2311.16452}, 2023.

\bibitem{openai_consequential_flag}
{OpenAI}.
\newblock Consequential flag.
\newblock \url{https://platform.openai.com/docs/actions/production/consequential-flag#consequential-flag}, 2024.

\bibitem{chatgpt_gpts_announcement}
OpenAI.
\newblock Introducing gpts.
\newblock \url{https://openai.com/blog/introducing-gpts}, 2024.

\bibitem{openai_chatgpt}
{OpenAI}.
\newblock Chatgpt.
\newblock \url{https://chatgpt.com/}, 2025.

\bibitem{openai_operator_2025}
{OpenAI}.
\newblock Introducing operator, 2025.

\bibitem{pawitan2025confidence}
Yudi Pawitan and Chris Holmes.
\newblock Confidence in the reasoning of large language models.
\newblock {\em Harvard Data Science Review}, 7(1), 2025.

\bibitem{peddinti2019reducing}
Sai~Teja Peddinti, Igor Bilogrevic, Nina Taft, Martin Pelikan, {\'U}lfar Erlingsson, Pauline Anthonysamy, and Giles Hogben.
\newblock Reducing permission requests in mobile apps.
\newblock In {\em Proceedings of the internet measurement conference}, pages 259--266, 2019.

\bibitem{prolific2024}
{Prolific}.
\newblock Prolific - recruitment platform for online research.
\newblock \url{https://www.prolific.com}.

\bibitem{razaghpanah2018apps}
Abbas Razaghpanah, Rishab Nithyanand, Narseo Vallina-Rodriguez, Srikanth Sundaresan, Mark Allman, Christian Kreibich, Phillipa Gill, et~al.
\newblock Apps, trackers, privacy, and regulators: A global study of the mobile tracking ecosystem.
\newblock In {\em The 25th Annual Network and Distributed System Security Symposium (NDSS 2018)}, 2018.

\bibitem{reis2000handbook}
Harry~T Reis and Charles~M Judd.
\newblock {\em Handbook of research methods in social and personality psychology}.
\newblock Cambridge University Press, 2000.

\bibitem{ringer2016audacious}
Talia Ringer, Dan Grossman, and Franziska Roesner.
\newblock Audacious: User-driven access control with unmodified operating systems.
\newblock In {\em Proceedings of the 2016 ACM SIGSAC Conference on Computer and Communications Security}, pages 204--216, 2016.

\bibitem{roesner2012user}
Franziska Roesner, Tadayoshi Kohno, Alexander Moshchuk, Bryan Parno, Helen~J Wang, and Crispin Cowan.
\newblock User-driven access control: Rethinking permission granting in modern operating systems.
\newblock In {\em 2012 IEEE Symposium on Security and Privacy}, pages 224--238. IEEE, 2012.

\bibitem{schwab2025makes}
Jasmin Schwab, Alexander Nussbaum, Anastasia Sergeeva, Florian Alt, and Verena Distler.
\newblock What makes phishing simulation campaigns (un) acceptable? a vignette experiment.
\newblock In {\em Network and Distributed System Security Symposium, NDSS 2025}, 2025.

\bibitem{seneviratne2015your}
Suranga Seneviratne, Aruna Seneviratne, Prasant Mohapatra, and Anirban Mahanti.
\newblock Your installed apps reveal your gender and more!
\newblock {\em ACM SIGMOBILE Mobile Computing and Communications Review}, 18(3):55--61, 2015.

\bibitem{shanmugarasa2021automated}
Yashothara Shanmugarasa, Hye-young Paik, Salil~S Kanhere, and Liming Zhu.
\newblock Automated privacy preferences for smart home data sharing using personal data stores.
\newblock {\em IEEE Security \& Privacy}, 20(1):12--22, 2021.

\bibitem{shao2022you}
Han Shao, Xiang Li, and Guodi Wang.
\newblock Are you tired? i am: Trying to understand privacy fatigue of social media users.
\newblock In {\em CHI Conference on Human Factors in Computing Systems Extended Abstracts}, pages 1--7, 2022.

\bibitem{sotirakopoulos2011challenges}
Andreas Sotirakopoulos, Kirstie Hawkey, and Konstantin Beznosov.
\newblock On the challenges in usable security lab studies: Lessons learned from replicating a study on ssl warnings.
\newblock In {\em Proceedings of the Seventh Symposium on usable Privacy and Security}, pages 1--18, 2011.

\bibitem{trimananda2022ovrseen}
Rahmadi Trimananda, Hieu Le, Hao Cui, Janice~Tran Ho, Anastasia Shuba, and Athina Markopoulou.
\newblock $\{$OVRseen$\}$: Auditing network traffic and privacy policies in oculus $\{$VR$\}$.
\newblock In {\em 31st USENIX security symposium (USENIX security 22)}, pages 3789--3806, 2022.

\bibitem{wang2024ali}
Han Wang, An~Zhang, Nguyen Duy~Tai, Jun Sun, Tat-Seng Chua, et~al.
\newblock Ali-agent: Assessing llms' alignment with human values via agent-based evaluation.
\newblock {\em Advances in Neural Information Processing Systems}, 37:99040--99088, 2024.

\bibitem{wiesinger2025agents}
Julia Wiesinger, Patrick Marlow, and Vladimir Vuskovic.
\newblock Agents, 2025.
\newblock \url{https://www.kaggle.com/whitepaper-agents}.

\bibitem{wijesekera2015android}
Primal Wijesekera, Arjun Baokar, Ashkan Hosseini, Serge Egelman, David Wagner, and Konstantin Beznosov.
\newblock Android permissions remystified: A field study on contextual integrity.
\newblock In {\em 24th USENIX Security Symposium (USENIX Security 15)}, pages 499--514, 2015.

\bibitem{jaff2024data}
Yuhao Wu, Evin Jaff, Ke~Yang, Ning Zhang, and Umar Iqbal.
\newblock An in-depth investigation of data collection in llm app ecosystems.
\newblock In {\em ACM Internet Measurement Conference (IMC)}, 2025.

\bibitem{wu2025isolategpt}
Yuhao Wu, Franziska Roesner, Tadayoshi Kohno, Ning Zhang, and Umar Iqbal.
\newblock {IsolateGPT: An Execution Isolation Architecture for LLM-Based Agentic Systems}.
\newblock In {\em Network and Distributed System Security (NDSS) Symposium}, 2025.

\bibitem{xie2014location}
Jierui Xie, Bart~Piet Knijnenburg, and Hongxia Jin.
\newblock Location sharing privacy preference: analysis and personalized recommendation.
\newblock In {\em Proceedings of the 19th international conference on Intelligent User Interfaces}, pages 189--198, 2014.

\bibitem{yao2023react}
Shunyu Yao, Jeffrey Zhao, Dian Yu, Nan Du, Izhak Shafran, Karthik Narasimhan, and Yuan Cao.
\newblock React: Synergizing reasoning and acting in language models.
\newblock In {\em International Conference on Learning Representations (ICLR)}, 2023.

\bibitem{yao2024survey}
Yifan Yao, Jinhao Duan, Kaidi Xu, Yuanfang Cai, Zhibo Sun, and Yue Zhang.
\newblock A survey on large language model (llm) security and privacy: The good, the bad, and the ugly.
\newblock {\em High-Confidence Computing}, page 100211, 2024.

\bibitem{zhan2023privacy}
Nicole Zhan, Stefan Sarkadi, and Jose Such.
\newblock Privacy-enhanced personal assistants based on dialogues and case similarity.
\newblock In {\em European Conference on Artificial Intelligence}. IOS Press, 2023.

\bibitem{zhan2022model}
Xiao Zhan, Stefan Sarkadi, Natalia Criado, and Jose Such.
\newblock A model for governing information sharing in smart assistants.
\newblock In {\em Proceedings of the 2022 AAAI/ACM Conference on AI, Ethics, and Society}, pages 845--855, 2022.

\bibitem{zhao2024llm}
Yanjie Zhao, Xinyi Hou, Shenao Wang, and Haoyu Wang.
\newblock Llm app store analysis: A vision and roadmap.
\newblock {\em ACM Transactions on Software Engineering and Methodology}, 2024.

\end{thebibliography}
